\documentclass[twocolumn]{aastex63}

\usepackage[english]{babel}
\usepackage{amsmath}
\usepackage[version=3]{mhchem}

\usepackage{xspace}
\usepackage{units}

\usepackage{color}

\newcommand{\ammo}{\ce{NH3}\xspace}
\newcommand{\ammostar}{\ce{NH3^*}\xspace}

\newcommand{\htwo}{\ce{H2}\xspace}
\newcommand{\ntwohp}{\ce{N2H+}\xspace}
\newcommand{\kms}{\ensuremath{\rm km\,s^{-1}}\xspace}
\newcommand{\kkms}{\ensuremath{\rm K\,km\,s^{-1}}\xspace}

\newcommand{\tk}{\ensuremath{T_{\rm k}}\xspace}
\newcommand{\tex}{\ensuremath{T_{\rm ex}}\xspace}
\newcommand{\sigv}{\ensuremath{\sigma_{\rm v}}\xspace}

\newcommand{\vlsr}{\ensuremath{V_{\rm LSR}}\xspace}
\newcommand{\cc}{\ensuremath{\rm cm^{-3}}\xspace}

\newcommand{\jybm}{\ensuremath{\rm Jy\,beam^{-1}}\xspace}
\newcommand{\MJysr}{\ensuremath{\rm MJy\,sr^{-1}}\xspace}

\newcommand{\meth}{\ensuremath{\rm CH_3OH}}

\newcommand{\sqcm}{\ensuremath{\rm cm^2}\xspace}
\newcommand{\persqcm}{\ensuremath{\rm cm^{-2}}\xspace}
\newcommand{\percc}{\rm cm^{-3}}
\newcommand{\pers}{\rm s^{-1}}
\newcommand{\perg}{\rm g^{-1}}
\newcommand{\mum}{\rm \mu m}
\newcommand{\ms}{\rm m\,s^{-1}}

\shortauthors{J.E. Pineda et al.}
\shorttitle{Ammonia Depletion in H-MM1}

\graphicspath{{./}{figs/}}

\received{}
\revised{}
\accepted{}

\begin{document}

\title{An Interferometric View of H-MM1.
I. Direct Observation of \ammo Depletion}

\correspondingauthor{Jaime E. Pineda}
\email{jpineda@mpe.mpg.de}

\author[0000-0002-3972-1978]{Jaime E. Pineda}
\affiliation{Max-Planck-Institut f\"ur extraterrestrische Physik, Giessenbachstrasse 1, D-85748 Garching, Germany}

\author[0000-0002-1189-9790]{Jorma Harju}
\affiliation{Department of Physics, P.O. Box 64, FI-00014, University of Helsinki, Finland}
\affiliation{Max-Planck-Institut f\"ur extraterrestrische Physik, Giessenbachstrasse 1, D-85748 Garching, Germany}

\author[0000-0003-1481-7911]{Paola Caselli}
\affiliation{Max-Planck-Institut f\"ur extraterrestrische Physik, Giessenbachstrasse 1, D-85748 Garching, Germany}

\author[0000-0002-9148-1625]{Olli Sipil\"a}
\affiliation{Max-Planck-Institut f\"ur extraterrestrische Physik, Giessenbachstrasse 1, D-85748 Garching, Germany}

\author[0000-0002-5809-4834]{Mika Juvela}
\affiliation{Department of Physics, P.O. Box 64, FI-00014, University of Helsinki, Finland}

\author[0000-0001-8211-6469]{Charlotte Vastel}
\affiliation{Universit\'e de Toulouse, UPS-OMP, IRAP, Toulouse, France}
\affiliation{CNRS, IRAP, 9 Avenue du Colonel Roche, BP 44346, F-31028 Toulouse Cedex 4, France}

\author[0000-0002-5204-2259]{Erik Rosolowsky}
\affiliation{Department of Physics, 4-181 CCIS, University of Alberta, Edmonton, AB T6G 2E1, Canada}

\author[0000-0001-6879-9822]{Andreas Burkert}
\affiliation{University Observatory Munich, Scheinerstr. 1, D-81679 Munich, Germany}
\affiliation{Max-Planck-Institut f\"ur extraterrestrische Physik, Giessenbachstrasse 1, D-85748 Garching, Germany}

\author[0000-0001-7594-8128]{Rachel K. Friesen}
\affiliation{Dunlap Institute for Astronomy and Astrophysics, University of Toronto, 50 St. George Street, Toronto M5S 3H4, Ontario, Canada}

\author{Yancy Shirley}
\affiliation{Steward Observatory, University of Arizona, 933 North Cherry Avenue, Tucson, AZ, 85721, USA}

\author[0000-0002-7026-8163]{Mar\'ia Jos\'e Maureira}
\affiliation{Max-Planck-Institut f\"ur extraterrestrische Physik, Giessenbachstrasse 1, D-85748 Garching, Germany}

\author[0000-0002-7497-2713]{Spandan Choudhury}
\affiliation{Max-Planck-Institut f\"ur extraterrestrische Physik, Giessenbachstrasse 1, D-85748 Garching, Germany}

\author[0000-0003-3172-6763]{Dominique M. Segura-Cox}
\altaffiliation{NSF Astronomy and Astrophysics Postdoctoral Fellow}
\affiliation{Max-Planck-Institut f\"ur extraterrestrische Physik, Giessenbachstrasse 1, D-85748 Garching, Germany}
\affiliation{Department of Astronomy, The University of Texas at Austin, 2500 Speedway, Austin, TX 78712, USA}

\author[0000-0002-1708-9289]{Rolf G\"usten}
\affiliation{Max-Planck-Institut f\"ur Radioastronomie, Auf dem H\"ugel 69, D-53121, Bonn, Germany}

\author[0000-0001-6004-875X]{Anna Punanova}
\affiliation{Ural Federal University, 620002, 19 Mira street, Yekaterinburg, Russia}

\author[0000-0002-9953-8593]{Luca Bizzocchi}
\affiliation{Dipartimento di Chimica ``Giacomo Ciamician'', Universit\`a di Bologna, Via F. Selmi 2, 40126 Bologna, Italy}
\affiliation{Max-Planck-Institut f\"ur extraterrestrische Physik, Giessenbachstrasse 1, D-85748 Garching, Germany}

\author[0000-0003-1312-0477]{Alyssa A. Goodman}
\affiliation{Harvard-Smithsonian Center for Astrophysics, 60 Garden Street, Cambridge MA 02138, USA}

\begin{abstract}
Spectral lines of ammonia, \ammo, are useful probes of the physical
conditions in dense molecular cloud cores. In addition to advantages
in spectroscopy, ammonia has also been suggested to be resistant to
freezing onto grain surfaces, which should make it a superior tool for
studying the interior parts of cold, dense cores. Here we present
high-resolution \ammo observations with the Very
Large Array (VLA) and Green Bank Telescope (GBT)
towards a prestellar core.
These observations show an outer region with 
a fractional \ammo abundance of 
$X(\ammo) = (1.975\pm0.005)\times 10^{-8}$ {($\pm$10\% systematic)}, 
but it also reveals that after all, the $X(\ammo)$
starts to decrease above a $\htwo$ column density of 
$\approx 2.6\times 10^{22}\,\persqcm$. 
We derive a density model for the core and 
find that the break-point in the fractional abundance occurs at the
density $n(\htwo)\sim 2\times 10^5\,\percc$, and beyond this point
the fractional abundance decreases with 
increasing density, following the power law $n^{-1.1}$. 
This power-law behavior is well reproduced by chemical models 
where adsorption onto grains dominates the 
removal of ammonia and
related species from the gas at high densities. We suggest that the
break-point density changes from core to core depending on the 
temperature and the grain properties, but that the depletion power law 
is anyway likely to be close to $n^{-1}$ owing to the dominance of
accretion in the central parts of starless cores.
\end{abstract}


\section{Introduction} \label{sec:intro}

Dense cores are the places where stars are formed. Their physical
conditions determine when the gravitational collapse starts and how
it proceeds \citep{diFrancesco_2007-PPV}. 
Observations of molecular lines allow us to extract information not only 
about the physical structure of dense cores but also about their
chemical composition \citep{Bergin_Tafalla_2007-Review}.

When high densities and low temperatures are present, gas-phase
molecules can be depleted because of accretion onto the dust grains. 
The effect of depletion is commonly observed in \ce{CO} (and {isotopologues}),
and it is identified as a strong decrease in the \ce{CO} abundance towards 
the central part of a dense core 
\citep[e.g.,][]{Caselli_1999-L1544_Depletion,Tafalla2004-Cores}.
Surprisingly, some nitrogen-bearing species, such as \ammo, \ntwohp and
\ce{CN}, do not present strong evidence for depletion 
\citep{Tafalla2004-Cores, Crapsi2007-L1544, 2008A&A...480L...5H,
2010A&A...513A..41H}. 
However, it is theoretically expected that all species, 
other than pure hydrogen species and helium, should be depleted by at 
least two orders of magnitude in the central 
region of a dense, starless core; this region can be called the zone of ``complete depletion'' \citep{Walmsley2004-L1544_Full_Depletion}.
In the past, modeling of single dish observations have provided indirect
evidence for depletion of \ntwohp in the L183 and L1544 dense cores
\citep{2005A&A...429..181P, Pagani2007-L183_Depletion_N2Hp, Redaelli2019-N2Dp_freezeout}.
Only recent observations with ALMA, resolving the densest 
parts of the prestellar core L1544, show evidence for depletion of
deuterated ammonia, \ce{NH2D} \citep{Caselli2022-L1544_NH3_Depletion}.

To validate the concept of complete depletion, it is important to 
investigate if also the common isotopolog of ammonia, \ammo, is also depleted 
at high densities. The starless core H-MM1 \citep{2004ApJ...611L..45J,
2011A&A...526A..31P} is an excellent target for this purpose.  
H-MM1 lies in relative isolation on the outskirts of the L1688 region in the nearby Ophiuchus molecular cloud. 
{The distance to L1688 is $\approx$138.4$\pm$2.6\,pc \citep{2018ApJ...869L..33O}}.
Ammonia line emission in the L1688 
region has been previously studied with single-dish observations
\citep{Friesen2017,Harju2017-HMM1_NH2D,Chen2019-Droplets,Auddy2019-Bfield,
Choudhury2020-Letter,Choudhury_2021-L1688_Map}. 
These studies show a clear transition from the supersonic turbulence to a
subsonic regime at the boundary of the H-MM1 core, and that the interior
parts of the core are characterized by a {high degree} of deuteration in
\ammo \citep{Harju2017-HMM1_NH2D}. This suggests that the densities and
temperatures in the central region are appropriate for \ammo depletion. 

High-resolution \ammo observations are also useful for the
interpretation of recent ALMA observations that revealed the presence
of a methanol (\meth) sheath around the dense core \citep{Harju2020-HMMS1}. Of the methanol
desorption mechanisms discussed in \cite{Harju2020-HMMS1}, shocks
and grain-grain collisions induced by vigorous turbulence or shear
instability should be associated with increased kinetic temperatures or
velocity dispersion in the methanol emission region. Ammonia provides
excellent means to determine these quantities and thereby test
these desorption scenarios.

Here we present \ammo (1,1) and (2,2) observations of the starless core
H-MM1 with the Karl G. Jansky Very Large Array (VLA), which we combine
with previous Robert C. Byrd Green Bank Telescope (GBT) observations
\citep{Friesen2017}. 
The acquired data allow us to study the \ammo emission at high angular 
resolution to resolve the possible \ammo depletion zone and to improve our knowledge of the origin of methanol.

\section{Data}

\subsection{VLA}
We conducted VLA observations of H-MM1 on 2020 January 9, 12, 14, and 18
in the D-array configuration (project 19B-178, PI: Pineda). 
We used the high-frequency K-band receiver and configured the WIDAR
correlator to observe the \ammo (1,1) and (2,2) lines with 8 and 4\,MHz
bandwidth windows, respectively. 
Each window has a channel separation of 5.208\,kHz ($\approx$0.065\,\kms).
The quasar 3C 286 is used as the flux calibrator, J1256-0547 as the
bandpass calibrator, and J1625-2527 as the phase calibrator. 

The calibration was performed using the VLA pipeline with the Common
Astronomy Software Applications package (CASA) version 5.6.2-2
\citep{McMullin2007-CASA}.  
This included Hanning-smoothing, and therefore the effective spectral resolution is 0.13\,\kms. 
The imaging used the \verb+tclean+ command with natural weighting and \verb+uvtaper+ 
to increase the signal to noise ratio, a restoring beam of 6\arcsec, 
and with multi-scale clean (with scales 0, 6, 18, and 54\arcsec 
and a \verb+smallscalebias+ parameter of 0).
We also included the \ammo single dish data obtained with the Robert C.
Byrd Green Bank Telescope \citep{Friesen2017} as a model image to recover
the extended emission 
{using the \texttt{startmodel} option in \texttt{tclean}.
This method initializes the clean model image as the 
single dish data.}
The final noise level achieved in the data cubes is
3.5\,m\jybm per channel, which corresponds to $\approx$0.21 K per channel.
{We estimate our absolute calibration uncertainty to be 10\%.}

\begin{figure*}
    \centering
    \includegraphics[height=6.71cm]{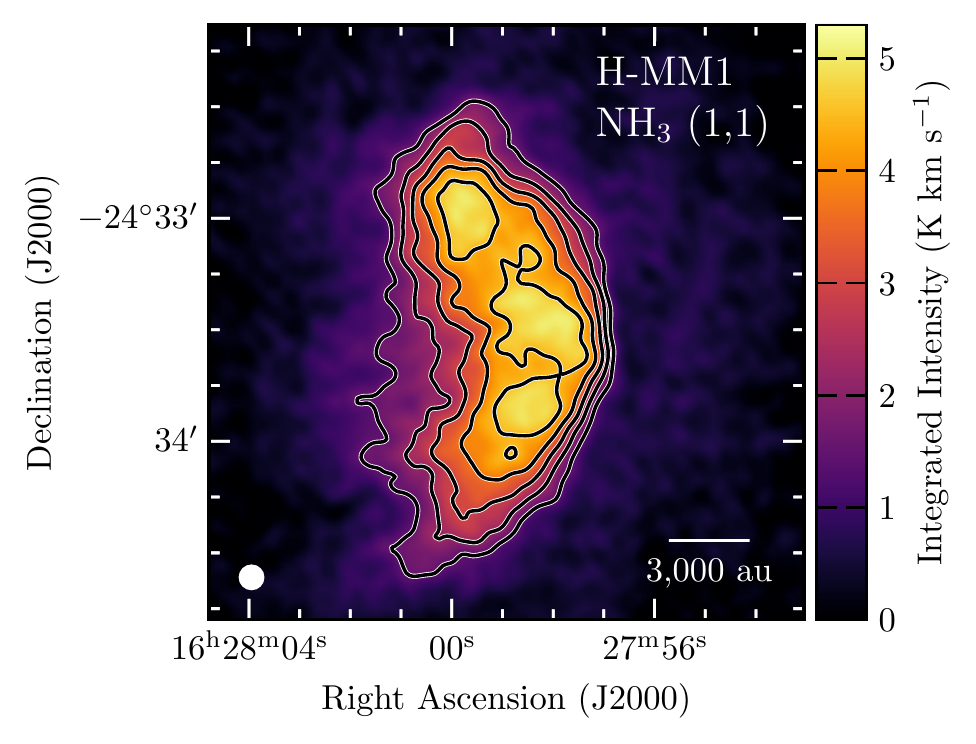}
    \includegraphics[height=6.71cm]{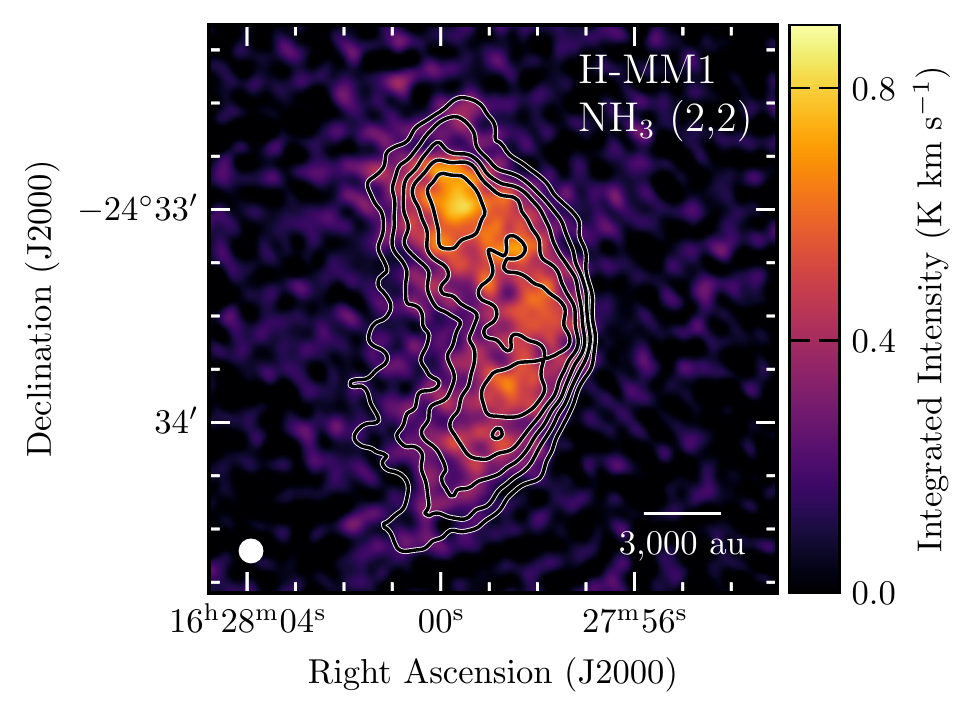}
    
    \caption{Integrated intensity maps of the \ammo (1,1) and (2,2) main
    line components are shown in {the} left and right panels, respectively.
    The contours are for \ammo (1,1) in both panels. The contour levels
    start from 1.5\,\kkms and the spacing is 0.75\,\kkms. The 
    beam size and the scale bar are shown at bottom left and right, respectively.}
    \label{fig:NH3_TdV}
\end{figure*}

The integrated intensity maps of the main hyperfine component 
of the \ammo (1,1) and (2,2) lines are shown in Fig.~\ref{fig:NH3_TdV}.

\subsection{Total {\htwo} Column Density}

\begin{figure}[htb]
    \centering
    \includegraphics[width=\columnwidth]{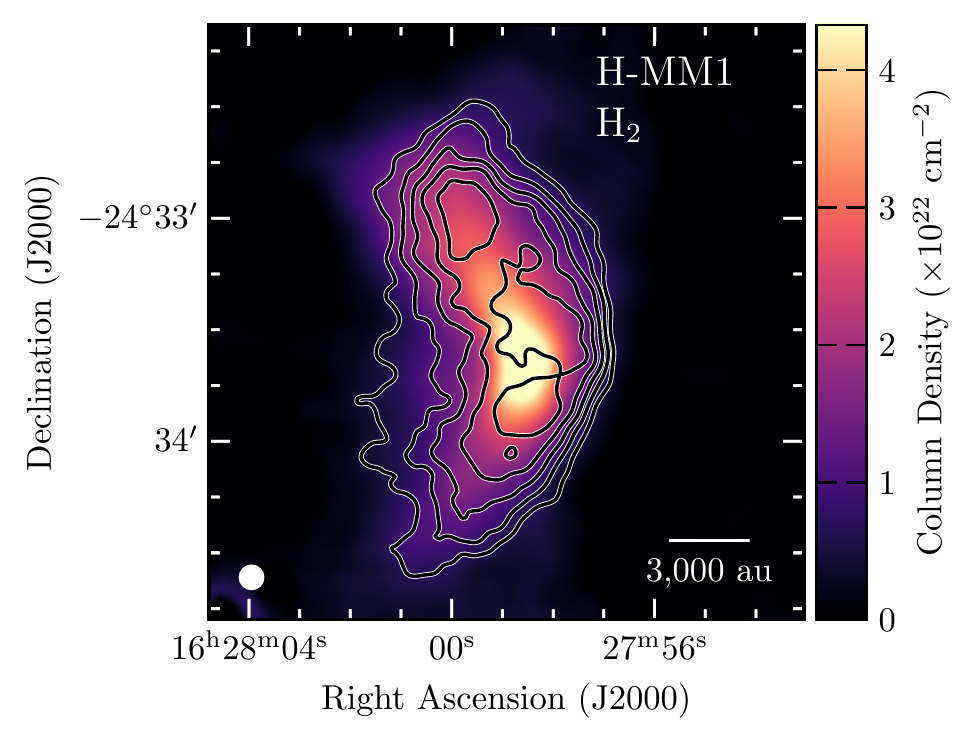}
    \caption{Dust derived total \htwo column density map of the H-MM1 region. 
    The map shows a peak close to the center of the image. 
    The resolution and scale bar are shown at bottom left and right, respectively.
    The contours show the \ammo (1,1) integrated intensity, as in Fig~\ref{fig:NH3_TdV}.
    \label{fig:column_maps_H2}}
\end{figure}

The H-MM1 core appears as an absorption feature in the 8 $\mum$ map
measured by the InfraRed Array Camera (IRAC) of the {\it Spitzer Space
Telescope}. 
We use this to derive a high-resolution $\htwo$ column
density map of the core. The method is described in Appendix \ref{sec:scatter} of
\cite{Harju2020-HMMS1}, and briefly summarized here. 
The method also uses
the 850 $\mum$ emission map from SCUBA-2 \citep{Pattle:2015bq}, and the
dust color temperature map, $T_{\rm C}$, derived from {\it Herschel}/SPIRE maps
\citep{Harju2017-HMM1_NH2D}. 
We assume the dust opacity model for unprocessed dust grains with thin ice mantles by
\cite{1994A&A...291..943O}. This model has a dust emissivity index of
$\beta=2.0$ over {\it Herschel}/SPIRE wavelengths (250--500 $\mum$).
The absorption cross-sections per unit mass of gas at 8 $\mum$ and
850\,$\mum$ are $\kappa_{8\,\mum}=8.86\, \sqcm\,\perg$ and
$\kappa_{850\,\mum}=1.13\times10^{-2}\, \sqcm\,\perg$,
respectively. The {\it Spitzer} map is smoothed to the resolution of the
VLA ammonia map, $6\,\arcsec$, before the derivation of the column
density map. The resulting $N(\htwo)$ distribution is shown in
Fig.~\ref{fig:column_maps_H2}. 

\cite{Lefevre2016} pointed out that scattering by dust particles may bias 
column density derived from $8\,\mum$ absorption. 
This issue is discussed in Appendix \ref{sec:scatter}. 
Simulations using different dust models
show that the effect of scattering is likely to be small compared to the
uncertainties related to the intensities of the foreground and background
emission components, and thermal dust emission at 8 $\mum$. 
The 8 $\mum$ background intensity in the region of H-MM1 is 
$\sim$10 $\MJysr$, whereas the scattered intensity according to our
simulations should stay below $0.1\,\MJysr$ (see Appendix \ref{sec:scatter}).

\section{Observational Results}
\subsection{Line Fits} \label{sec:linefit}

\begin{figure*}[th]
    \centering
    \includegraphics[height=6.595cm]{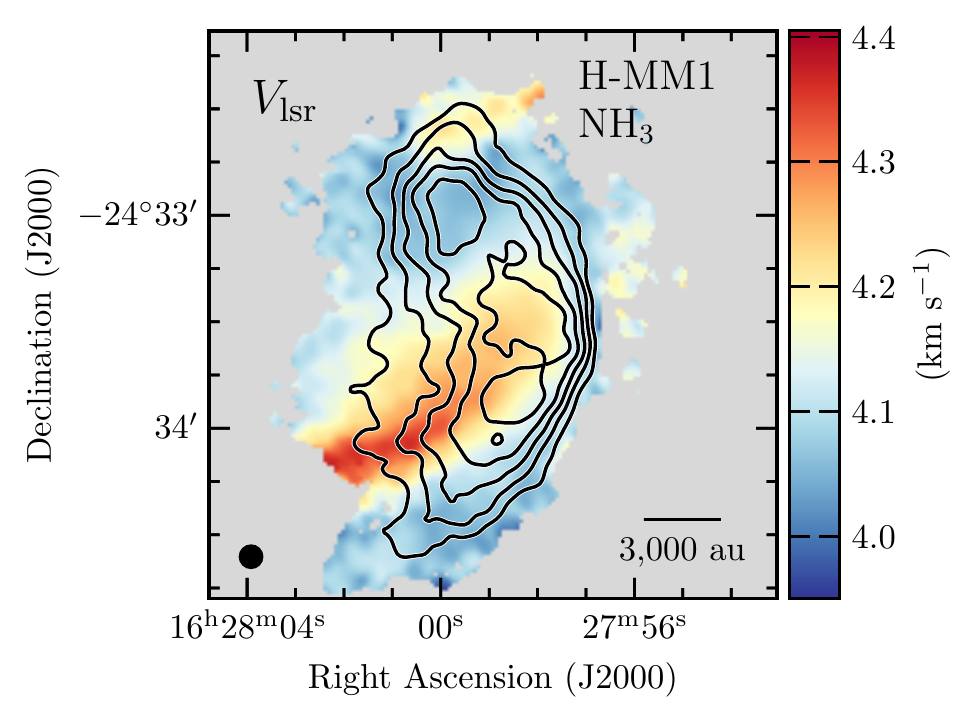}
    \includegraphics[height=6.595cm]{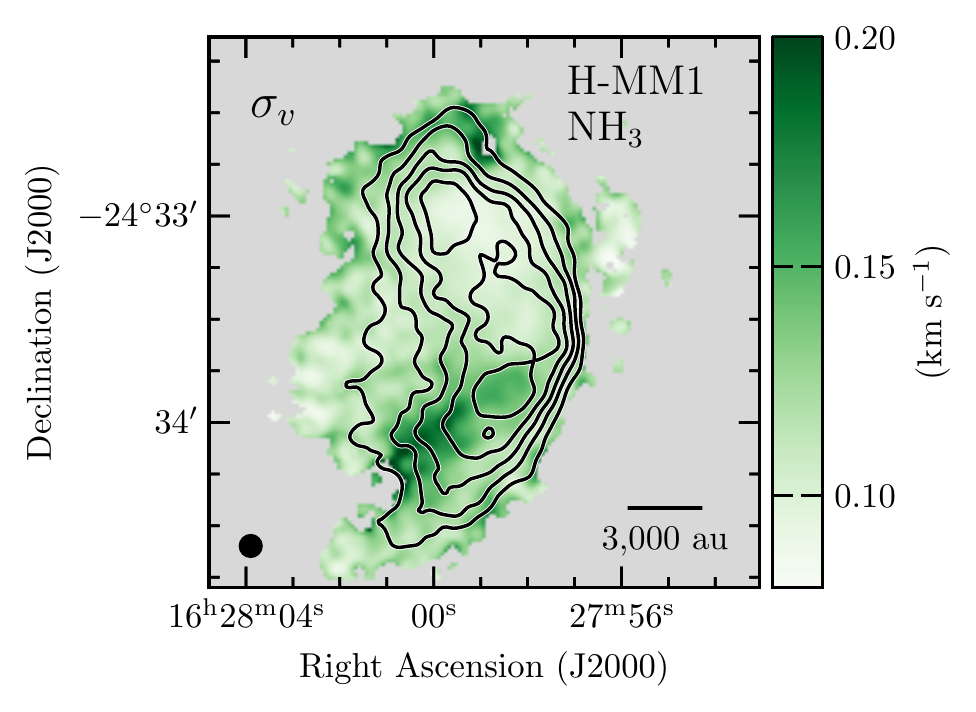}
    \caption{Centroid velocity and velocity dispersion maps derived from \ammo are shown in {the} 
    left and right panels, respectively. 
    The velocity fields show an elongated redshifted feature connecting the south-east edge and the
    central region of the core.
    The contours and the markers are as in Fig~\ref{fig:NH3_TdV}.
    \label{fig:NH3_velo}}
\end{figure*}

\begin{figure}[thb]
    \centering
    \includegraphics[width=\columnwidth]{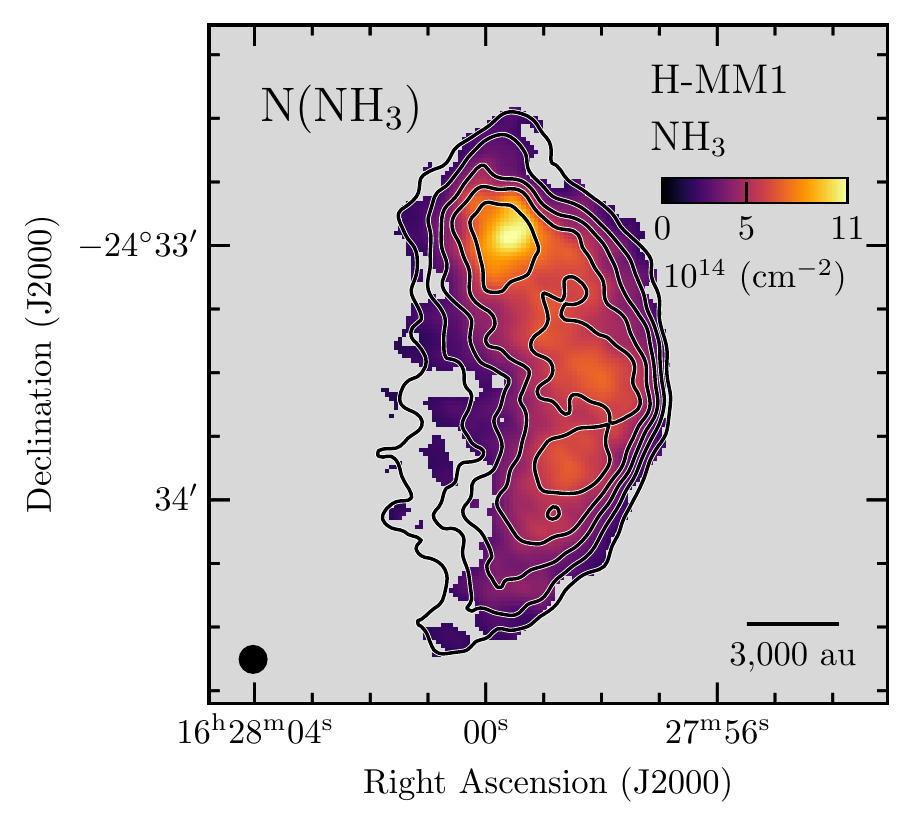}
    \caption{\ammo column density map.
    The central part of the map (where \htwo peaks)
    shows a nearly constant \ammo column density,
    while the peak is located in the northern part of the core. 
    The contours and the markers as in Fig~\ref{fig:NH3_TdV}.
    \label{fig:column_maps}}
\end{figure}

\begin{figure*}[hb]
    \centering
    \includegraphics[height=7.75cm]{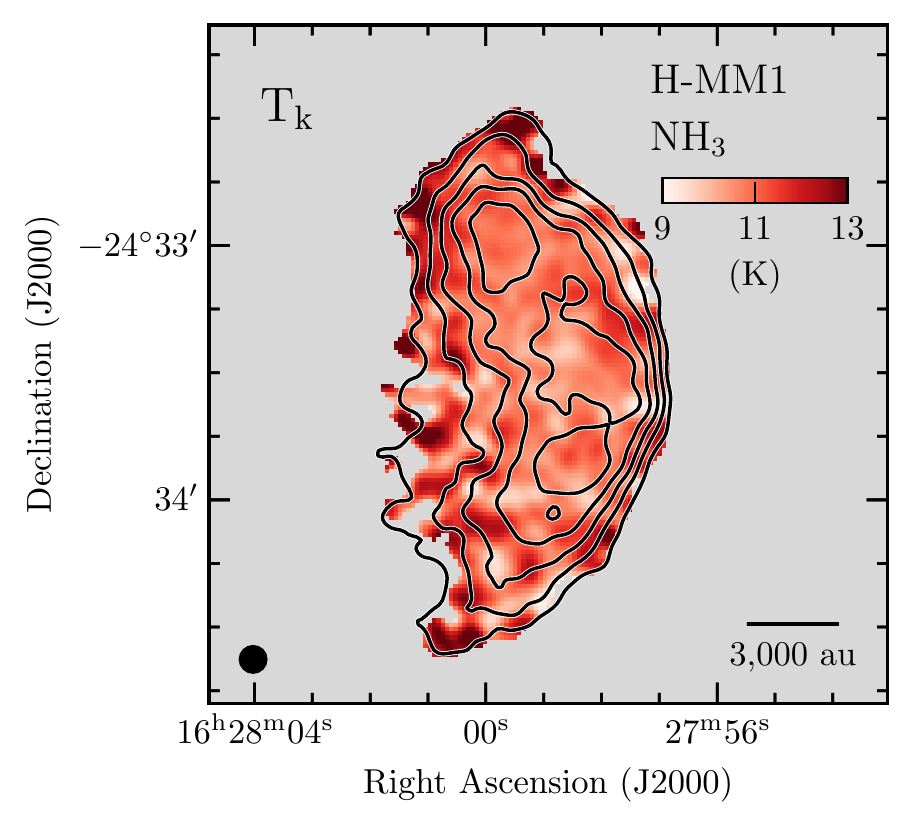}
    \includegraphics[height=7.75cm]{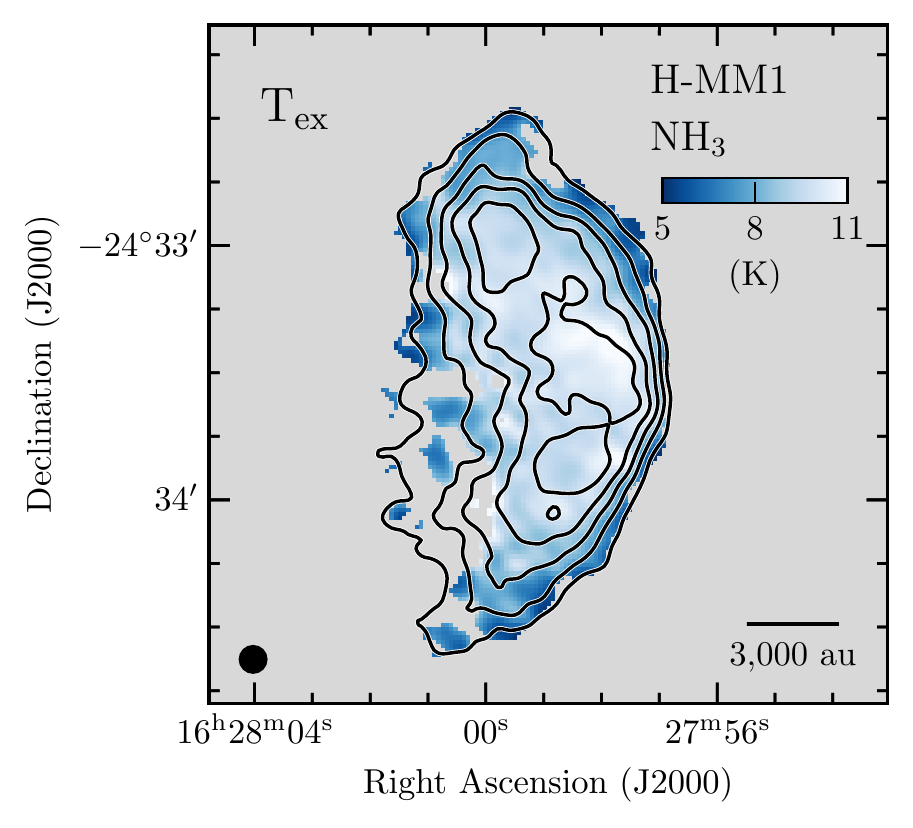}
    \caption{Kinetic and excitation temperature maps for H-MM1 are shown in {the} 
    left and right panels, respectively.
    The contours and the markers are as in Fig~\ref{fig:NH3_TdV}.
    These maps only include pixels where 
    the uncertainty is smaller than 1\,K. \label{fig:TexTk}}
\end{figure*}

We simultaneously fit the \ammo(1,1) and (2,2) line profiles with \verb+pyspeckit+ using the \verb+cold-ammonia+ model \citep{Friesen2017}, which gives a centroid velocity, \vlsr, velocity dispersion, \sigv, kinetic temperature, \tk, excitation temperature, \tex, and total column density of \ammo, $N(\ammo)$, with an assumed ortho-to-para ratio of 1 \footnote{If the column density of para-\ammo is desired then a correction factor of 0.5 on the \ammo column densities and fractional abundances should be applied}. 
This model assumes {that
only} the rotational levels (1,1) and (2,2) are populated, 
which is appropriate for the low temperatures seen in this region.
We adopt a number of criteria to ensure that the different fit results are accurate; 
this approach is similar to what was used in \cite{Pineda2021-Ion_Neutrals}.
The mean value of the kinetic temperature (\tk) in pixels with 
uncertainty less than 1\,K is 11\,K. 
For pixels with uncertainties in \tk larger than 1\,K we re-run the fit, but with a fixed \tk of 11\,K.
The kinematic parameters, \vlsr and \sigv, are usually well determined even when the 
kinetic and excitation temperatures are poorly constrained, thanks to the many hyperfine components. 
We discard all velocity determinations if the uncertainty in \vlsr or \sigv is larger 
than 0.02\,\kms and 0.015\,\kms, respectively.
We consider that the \tex and \tk are well determined only when their derived uncertainties are smaller than 1\,K, 
and that the column density is well constrained where \tex and \tk are well constrained.
The centroid velocity and the velocity dispersion maps are shown in Fig.~\ref{fig:NH3_velo}, 
the \ammo column density is shown in Fig.~\ref{fig:column_maps}, 
and the excitation and kinetic temperature maps are shown in Fig.~\ref{fig:TexTk}.
The results from the \ammo line fits are discussed below. 

\subsection{Radial Velocity and Velocity Dispersion}
The velocity
centroids (Fig.~\ref{fig:NH3_velo}-left) range from 
$\approx$4.0\,\kms to $\approx$4.4\,\kms across the core. 
Prominent in this distribution is a narrow feature at velocities $>$4.3\,\kms that seems 
to connect the South-East end and the center of the core. 
The velocity dispersion (Fig.~\ref{fig:NH3_velo}-right) is small across the map.
The narrowest line, $\approx$0.09\,\kms, is found in the northern part of the core, whereas the largest
values, $>$0.15\,\kms, lie in the southeast, in the region of the highest \vlsr  values. 
The locations of large \sigv 
form a narrow strip, but the orientation of this 
strip deviates from that of the red-shifted $V_{\rm LSR}$ feature.

\subsection{Kinetic Temperature and Excitation Temperature}

\begin{figure*}[htb]
    \centering
    \includegraphics[width=0.495\textwidth]{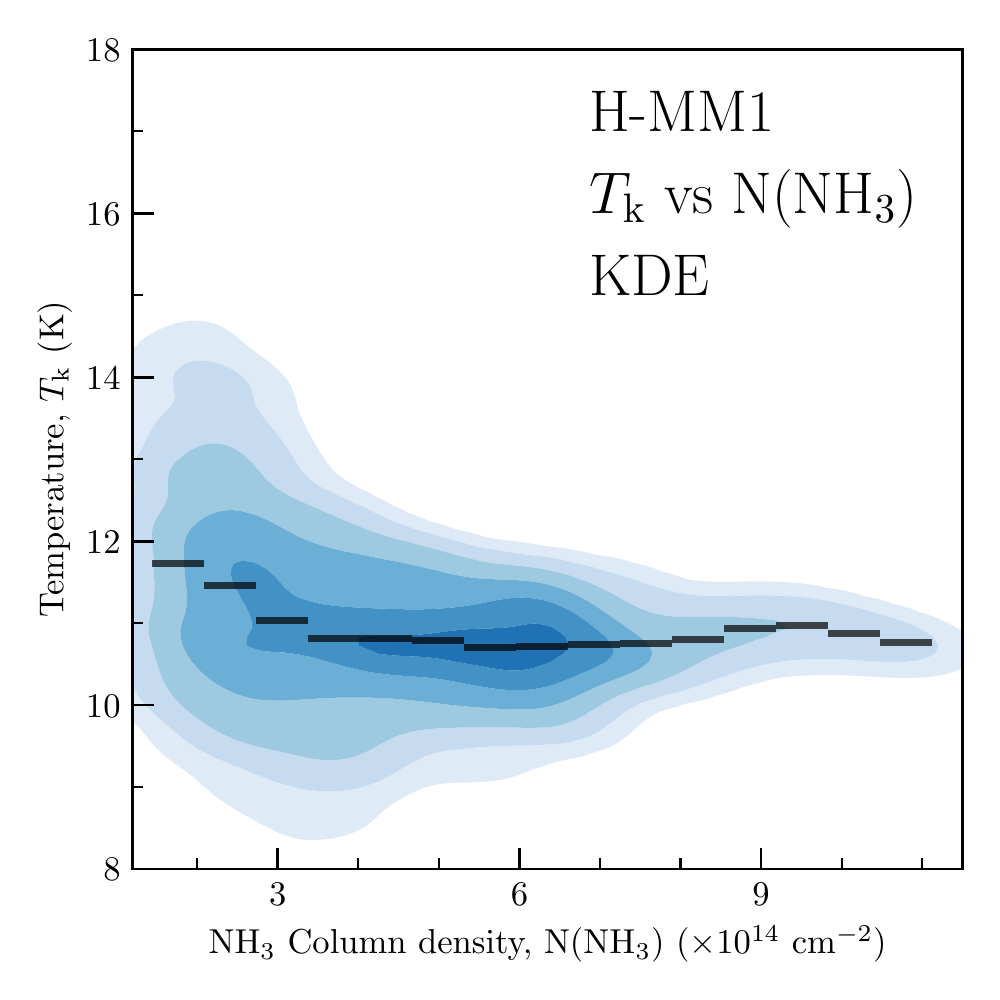}
    \includegraphics[width=0.495\textwidth]{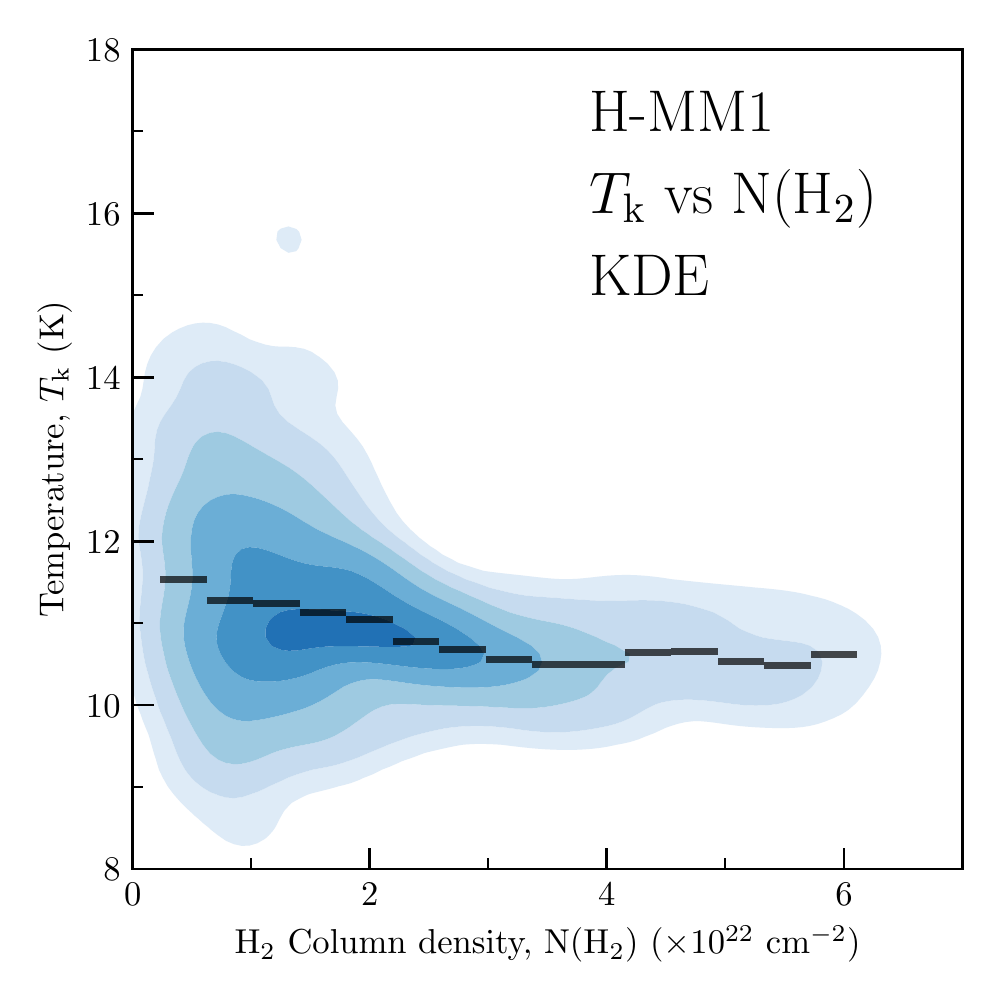}
    \caption{The Kernel Density Estimates (KDE) of 
    the kinetic temperature as a function of \ammo and \htwo 
    column density are shown in {the} left and right panels, respectively. 
    The black horizontal segments show the mean values within the column density bins. The gas temperature drops from $\approx$12 K down to
    $\approx11$ K towards the highest column densities in the core. 
    \label{fig:Tk_vs_N_NH3}
    }
\end{figure*}

The \tk distribution is almost flat in the central region, 
but it shows a tendency to increase toward the core boundaries (Fig.~\ref{fig:TexTk}).  
The average temperature in the 
core center is 11\,K, and it rises to $\approx$12\,K at lower $\ammo$ and $\htwo$ column densities (Fig.~\ref{fig:Tk_vs_N_NH3}). 
The \tex of the $\ammo(1,1)$ line varies from $\approx$5\,K at the core boundaries to 
$\approx$11\,K in the central regions. 
This indicates that the line is thermalized in the densest parts of the core {(see also Appendix~\ref{sec:Tex_Tk_H2})}.

\subsection{Ammonia Abundance\label{sec:X_calc}}
\begin{figure*}[htb]
    \centering
    \includegraphics[height=7.75cm]{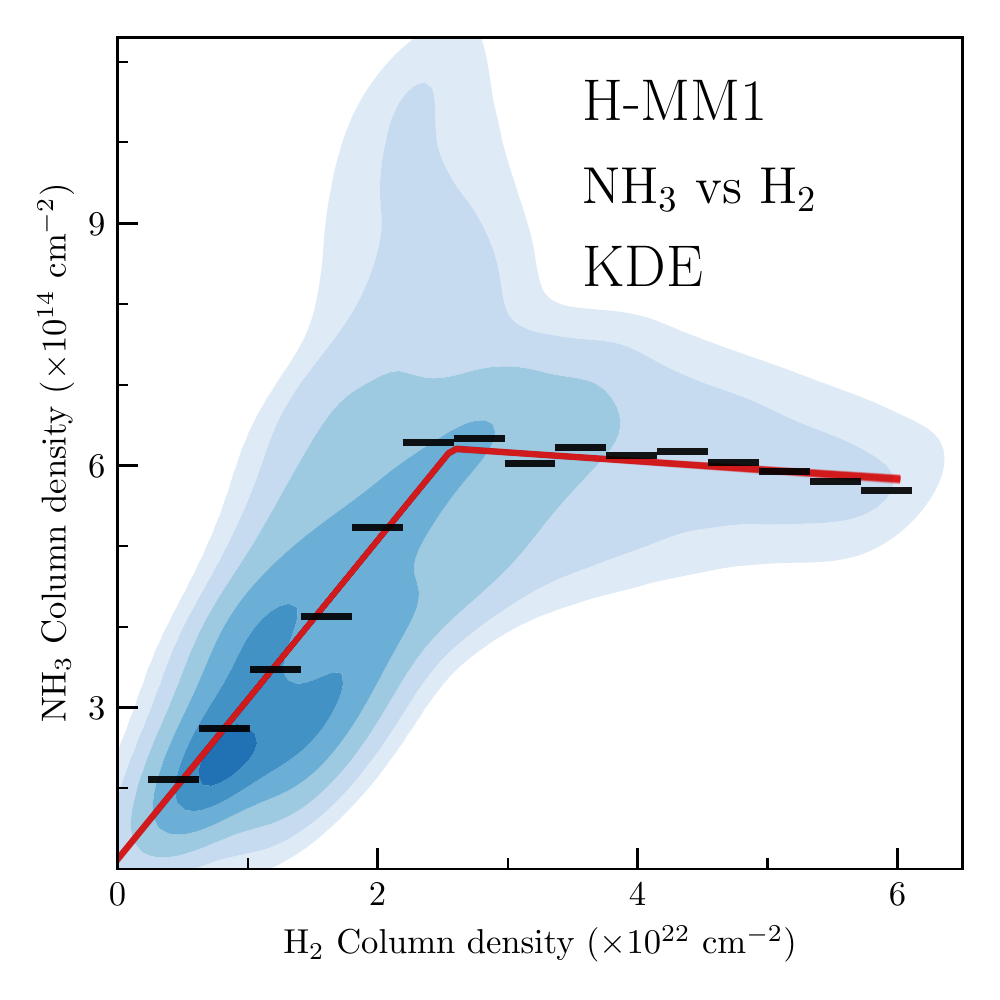}
    \includegraphics[height=7.75cm]{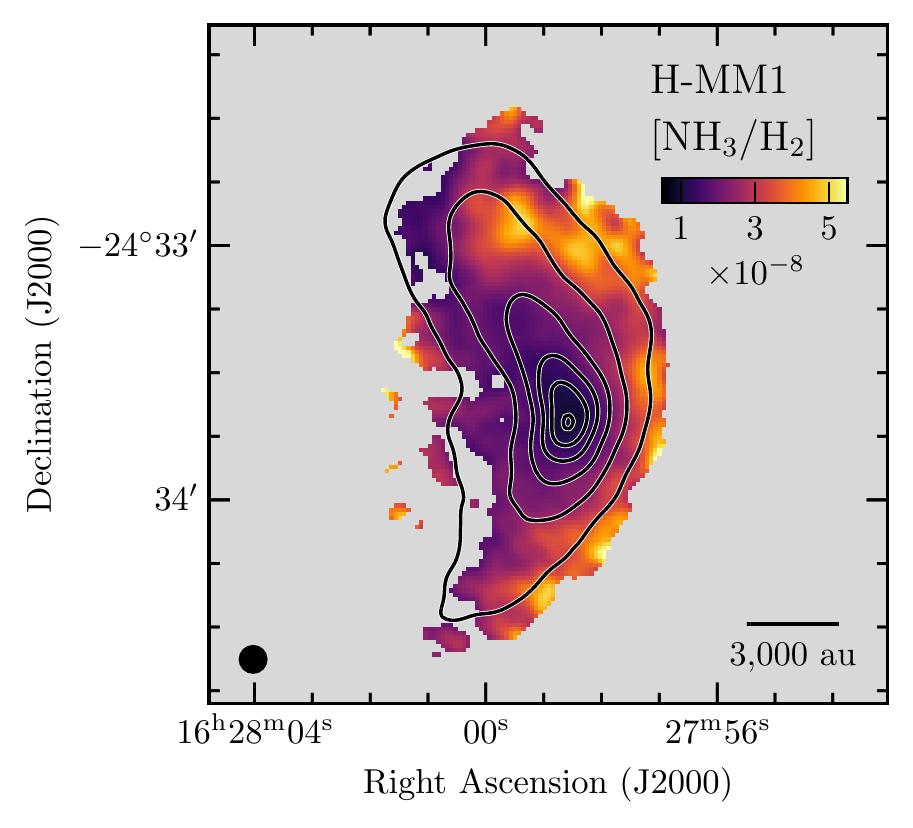}
    \caption{Depletion of \ammo in the center of H-MM1.
    {\it Left:} KDE comparison between \ammo and \htwo column densities. 
    This presents a positive correlation below $N(\htwo)<2.6\times
    10^{22}$ cm$^{-2}$, whereas for higher $N(\htwo)$ values the \ammo
    column density is nearly constant or decreasing to a small degree. 
    The black horizontal segments show the mean values within the 
    $N(\htwo)$ bins. The red line shows a broken straight line fit to 
    the data (see text).
    {\it Right:} Map of the fractional \ammo abundance,
    $X(\ammo)=[\ammo]/[\htwo]$, calculated as the ratio between the
    column densities. The \ammo abundance drops significantly in the
    central region, and the distribution is asymmetric so that 
    ammonia is more abundant on the Western side of the core as
    compared to the Eastern side. 
    The contours {show the \htwo column density 
    starting at $1\times 10^{22}$ \persqcm 
    and with a spacing of $1\times 10^{22}$ \persqcm.
    The beam size and scale bar are shown}
    as in Fig~\ref{fig:NH3_TdV}.
    \label{fig:NH3_abundance}}
\end{figure*}

An important quantity is the molecular fractional abundance of \ammo with respect {to} \htwo,
$X(\ammo) = [\ammo]/[\htwo]$.
This quantity is key to accurately determine total {\htwo} column densities from
\ammo observations {when no \htwo column density map is
available}, 
while it is also important to constrain chemical models \citep[e.g.,][]{Sipila2019-NH3_no_freezeout}.  
The $N(\ammo)$ versus $N(\htwo)$ relation is expected to be linear 
where $X(\ammo)$ is constant.  

The comparison of all pixels with an accurate \ammo and \htwo column density determinations is presented in Fig.~\ref{fig:NH3_abundance}-left.
It shows a region with a linear rise followed by a sharp change to a nearly constant $N(\ammo)$ as the $\htwo$ column density increases. 
We have fitted a broken straight line of the form 
\begin{equation}
    N(\ammo)[x] = 
    \begin{cases}
    a\cdot x + b & x<x_0 \\
    c\cdot x + (a-c)\cdot x_0 + b & x\ge x_0
    \end{cases}
\end{equation}
to the data points using \verb+emcee+ \citep{EMCEE}.
The results are listed in Table~\ref{tab:X_NH3_fit} and 
shown by the red curve in Fig.~\ref{fig:NH3_abundance}-left, 
while a more detailed 
description of the fit is given in Appendix~\ref{sec:X_fit}.

\begin{deluxetable}{ccc}
\tablecaption{Summary of Linear fit \label{tab:X_NH3_fit}}
\tablewidth{0pt}
\tablehead{
\colhead{Parameter} & \colhead{Value} & \colhead{Unit} 
}
\startdata
$a$ & $1.975_{-0.005}^{0.005}$ & $\times10^{-8}$\\
$b$ & $1.121_{-0.008}^{0.008}$ & $\times10^{14}$ cm$^{-2}$\\
$x_0$ & $2.575_{-0.004}^{0.003}$ & $\times10^{22}$ cm$^{-2}$\\
$c$ & $-0.110_{-0.005}^{0.005}$ & $\times10^{-8}$ \\
\hline
\enddata
{\tablecomments{Due to the systematic calibration uncertainties of the \ammo observations, 
we list a 10\% systematic uncertainty to $a$ and $c$ in addition to the statistical one in the text.}}
\end{deluxetable}

A different approach to estimate the molecular fractional abundance is 
to determine the fractional abundance at every pixel by 
taking the ratio of both column density estimates (see the map
in Fig.~\ref{fig:NH3_abundance}-right). 
Here we see a higher $X(\ammo)$ 
on the outskirts of the dense core, while in the central region there is
a sharp fall-off in the determined fractional abundance. 
Both the decreasing tendency in the $N(\ammo)$ versus $N(\htwo)$
correlation at high column densities in Fig.~\ref{fig:NH3_abundance}-left
and the minimum in the fractional abundance seen in
Fig.~\ref{fig:NH3_abundance}-right indicate that ammonia is strongly
depleted in the center of the core. 

\section{Discussion}

\subsection{Gas Kinematics}

The velocity dispersion, \sigv, derived from \ammo
observations changes only little within the core and indicates subsonic
non-thermal motions.  The H-MM1 core is thus a ``coherent core'' as
termed by \cite{1998ApJ...504..207B} and \cite{Goodman_1998-coherence} 
\citep[see also][]{Pineda:2010jq}. The kinematics of H-MM1 and other coherent 
cores in the Ophiuchus region have been previously studied using observations 
with the Green Bank Telescope (GBT) by \cite{Auddy2019-Bfield}, 
\cite{Chen2019-Droplets}, and \cite{Choudhury2020-Letter}. 
The thermal r.m.s. 
velocity of \ammo molecules at 11\,K is $73\,\ms$ in one dimension. The 
broadening owing to the channel width of the Hanning smoothed spectra, 
$130\,\ms$, corresponds to an instrumental $\sigma$ of $52\,\ms$ 
(see Appendix A of \citealt{Choudhury2020-Letter}). So, the median 
velocity dispersion, 
$\sigv \sim 110\,\ms$, from the present VLA observations indicates 
a median non-thermal dispersion of $\sigma_{\rm NT} \sim 63\,\ms$, 
or a median sonic Mach number of 
$\sim 0.3$. This value is approximately half of that derived by \cite{Choudhury2020-Letter} 
for the narrow $\ammo$ line component from GBT observations with an angular 
resolution of $31\arcsec$. 

\begin{figure}[ht]
    \centering
    \includegraphics[width=\columnwidth]{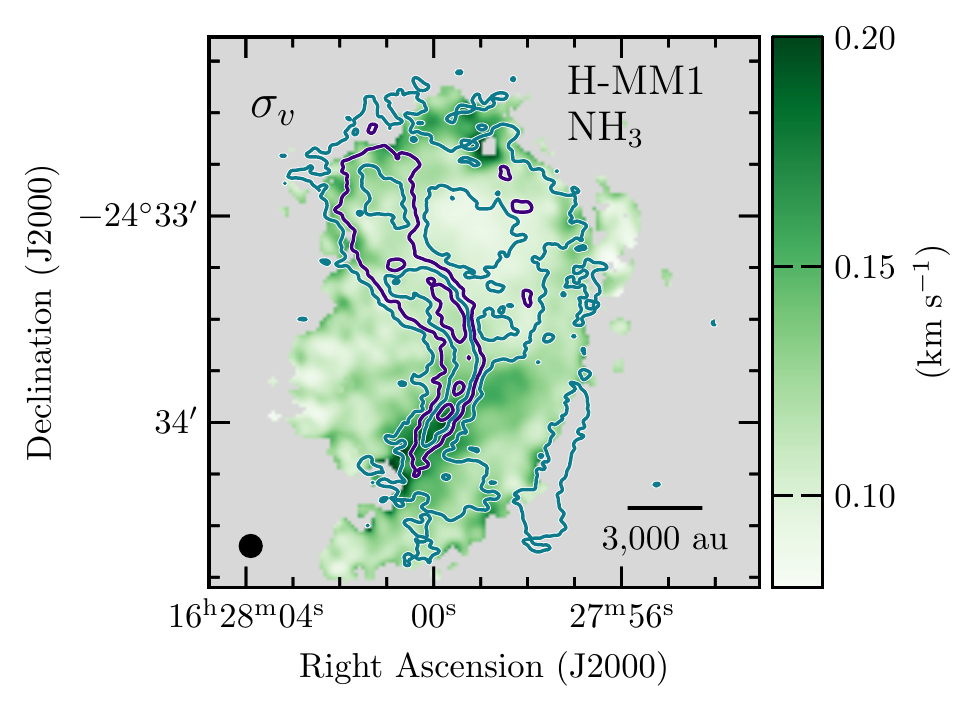}
    \caption{Velocity dispersion of $\ammo$ lines (color map) and the distribution of bright 96.7\,GHz methanol emission (contours) in H-MM1. The contour levels go from 0.8 to 4.0 K\,$\kms$ by steps of 0.8\,K\,$\kms$ {and alternating between
    blue and purple}.     \label{fig:sigv+meth}}
\end{figure}

The velocity dispersion attains values clearly higher than the median
in a narrow region in the southeastern part of the core. This region
overlaps partially with the southern ``tail'' of the bright $\meth$
emission mapped by \cite{Harju2020-HMMS1}. The integrated intensity contours of four methanol lines at 96.7 GHz are overlaid onto the $\ammo$ \sigv map in Fig.~\ref{fig:sigv+meth}. 
However, as can be seen in Fig.~\ref{fig:sigv+meth}, methanol emission 
arises mostly from regions where the dispersion is close to the average. 
Therefore, the present data do not support the idea that efficient $\meth$ desorption is only associated with enhanced turbulence or large velocity fluctuations, lending credence to the chemical desorption scenario of methanol formation  at the edge of the catastrophic CO {freeze-out zone} \citep{Vasyunin2017-COMs}. 
Perhaps it is worth to notice, though, that the most quiescent region in the northern part of the core with
$\sigv \sim 90\,\ms$ (implying $\sigma_{\rm NT} < 10\,\ms$) is
clearly devoid of $\meth$.

The elongated redshifted feature seen in the velocity centroid map at  $\vlsr > 4.3\,\kms$ seems to connect the compact dense core with the outer regions.
This feature bears resemblance to the streamer detected using gas tracers around binary protostar Per-emb-2 \citep{Pineda2020-Streamer}. 
We will discuss the core kinematics and the nature of this possible streamer in detail in a separate paper.

\subsection{Gas Temperature}

The gas temperature shows only a small drop toward the core
center. The temperature 11\,K in the inner parts is consistent with
previous gas and dust temperature determinations using GBT and
{\it Herschel} data 
\citep{Harju2017-HMM1_NH2D,Auddy2019-Bfield,Choudhury2020-Letter}. 
The temperature profile in
H-MM1 is very different from that in the well-studied prestellar core
L1544, where the central temperature reaches down to $\approx 6$\,K
\citep[][]{Crapsi2007-L1544}, 
or the starless core CB17, where the central temperature 
reaches down to $\approx 9.5$\,K \citep{Spear2021-CB17}.
The main difference is that H-MM1 is located in
an environment with strong radiation field  \citep[e.g.,][]{2013MNRAS.428.2617R}. 
Far-infrared radiation which penetrates the core is
absorbed by dust particles, and this heating is conveyed to gas by
gas-dust collisions. The density in the core, in excess of
$10^5\,\percc$, is high enough to make gas-dust thermal coupling
efficient \citep{2001ApJ...557..736G,2002A&A...394..275G,2019ApJ...884..176I}.

\subsection{Ammonia Depletion}\label{ss:depletion}

\begin{figure}[ht]
    \centering
    \includegraphics[width=\columnwidth]{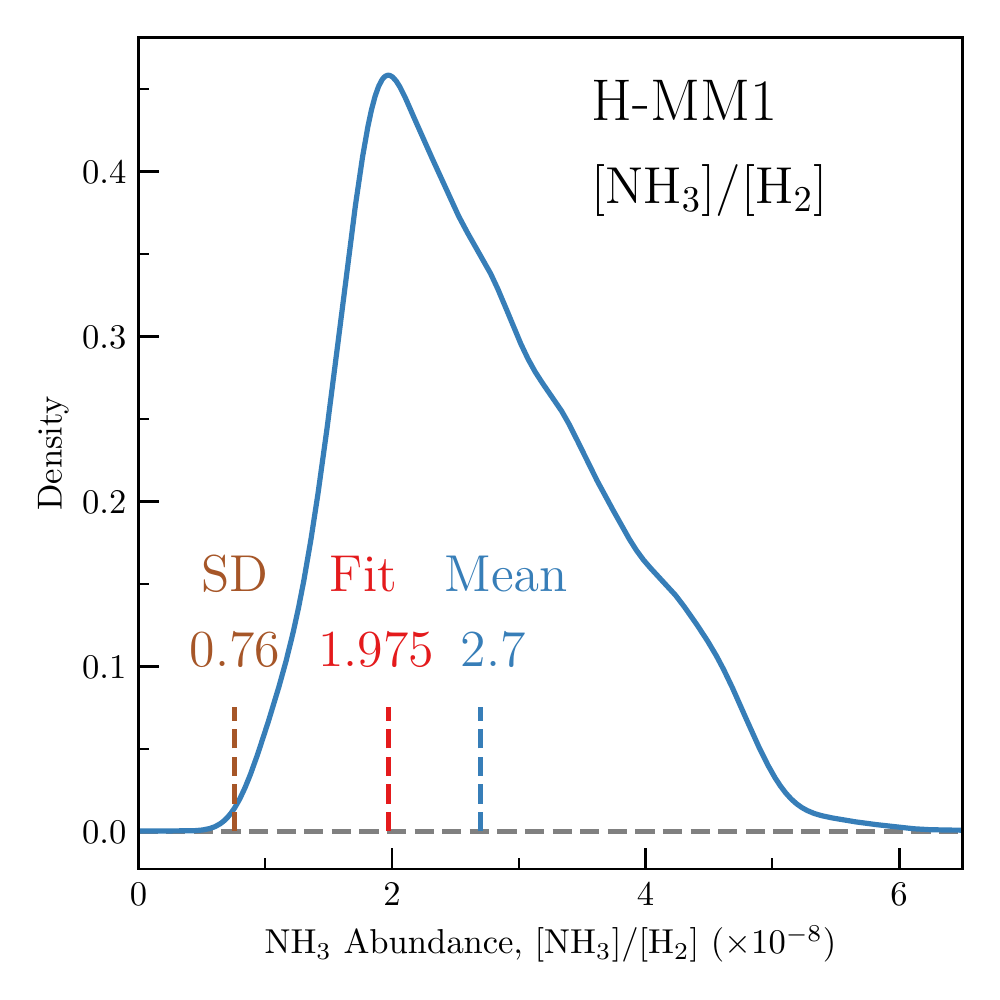}
    \caption{{Probability density function estimated with} 
    KDE of the fractional \ammo abundance towards H-MM1, $[\ammo]/[\htwo]$.
    The value estimated from single dish observations \citep{Harju2017-HMM1_NH2D} 
    is shown by the vertical brown line, $[\ammo]/[\htwo]_{SD}=7.6\times10^{-9}$. The mean value of the column density ratio is marked with a 
    vertical blue line, $[\ammo]/[\htwo]_{int}=2.7\times10^{-8}$,
    and the slope of the linear fit is marked with a red line,
    $[\ammo]/[\htwo]_{fit}=
    {(}1.975{\pm0.005)}\times 10^{-8}$ 
    {($\pm$10\% systematic)}.
    \label{fig:X_NH3_KDE}}
\end{figure}

Figure~\ref{fig:X_NH3_KDE} shows the Kernel Density Estimates (KDE) 
of the fractional \ammo abundance derived.
The mean  fractional \ammo abundance, $[\ammo]/[\htwo]_{int}$, derived as the ratio between 
$N(\ammo)$ and $N(\htwo)$ is $2.7\times10^{-8}$ (blue marker in Fig.~\ref{fig:X_NH3_KDE}), 
which is comparable to the values derived in other regions.  
However, this abundance value is 
higher by a factor of 3.5 compared to that derived from the single dish 
observations ($[\ammo]/[\htwo]_{SD} = 7.6\times10^{-9}$, \citealt{Harju2017-HMM1_NH2D}), 
which is shown with the brown marker in Fig.~\ref{fig:X_NH3_KDE}.
The fractional \ammo abundance derived from the linear fit 
($[\ammo]/[\htwo]_{fit}={(}1.975{\pm0.005)}\times 10^{-8}${, $\pm$10\% systematic})
is shown by the red marker in Fig.~\ref{fig:X_NH3_KDE}. 
The fractional \ammo abundance derived for H-MM1 
using the interferometric data is comparable 
to those derived from 
modelling single dish observations of L1498 and L1517 \citep[][]{Tafalla2004-Cores},
see Table~\ref{tab:X_NH3}.
This important difference between the interferometric and single dish results 
is due to the size of the depletion region, which is comparable 
to the single dish beam size. 
This highlights the crucial importance of 
deep spatially resolved maps of dense cores to study the depletion regions.

The fact that $\ammo$ is depleted in the central parts of the core is
evident from Fig.~\ref{fig:NH3_abundance}. The failure of
positive correlation between $N(\ammo)$ and $N(\htwo)$ at high column
densities (Fig.~\ref{fig:NH3_abundance}-left), and the deficiency of
ammonia in the core center (Fig.~\ref{fig:NH3_abundance}-right), cannot
be explained by extremely high optical thicknesses. The optical thicknesses of
the four satellite groups of \ammo (1,1) are clearly below 1 even toward the
center of the core, implying that the drop in the derived ammonia abundance is 
not an artefact caused by radiative transfer effects.

According to the two-slope fit presented in Sect.~\ref{sec:X_calc}, the break-point
until which the fractional ammonia abundance is $1.975\times10^{-8}$,
occurs at the column density $N(\htwo)=2.575\times10^{22}\,\persqcm$.
We constructed a three-dimensional model of the core by fitting a
Plummer-type function to its horizontal column density profiles 
(cuts parallel {to} the right ascension axis). 
{The inversion method is adopted from
\cite{2011A&A...529L...6A}, and it is explained 
in more detail in Appendix~{\ref{sec:density_model}.}
Using column densities derived
from $8\,\mum$ absorption, the core model does not include the ambient
cloud visible in the {\it Herschel}/SPIRE maps of this region. We assume
that the offset, $N(\ammo)\sim10^{14}\,\persqcm$, in the $N(\ammo)$ vs.
$N(\htwo)$ correlation shown in Fig.~\ref{fig:NH3_abundance} is caused by
ammonia residing in this ambient cloud.  
The peak
density of the core model is $n(\htwo)=1.3\times10^{6}\,\percc$. 
The relationship between the densities and column densities in this
model {is} shown in Fig.~\ref{fig:col2dens}. In this correlation, we
only include cells that have densities higher than
$5\times10^4\,\percc$. 
{The choice of the threshold density is based on the high
excitation temperatures of the {\ce{NH3}} lines ($>8$\,K, Fig.\,5-right),
which indicate that the emission is dominated by high-density gas.
According to the excitation calculations presented in 
\cite{Shirley2015-molecules} (their Fig.\,2) for
gas at $T_{\rm kin}=10$\,K, the $T_{\rm ex}$ of the $\ammo(1,1)$
transition exceeds 8\,K at the density $n(\htwo)\sim
5\times10^4\,\percc$. Using a lower density threshold would decrease
the mean densities but leave the power law, ${\bar n} \propto N^{0.8}$, unchanged.}

Line segments through the core having a column density around
$N(\htwo)=2.6\times10^{22}\,\persqcm$ where $X(\ammo)$ turns
down, have an average density of $(2.1\pm0.1)\times10^{5}\,\percc$.
One can therefore assume that the depletion of $\ammo$ starts to take
effect at this density, whereas at lower densities the fractional
ammonia abundance is nearly constant, $2\times10^{-8}$.

\begin{figure*}[htb]
    \centering
    \includegraphics[height=6cm]{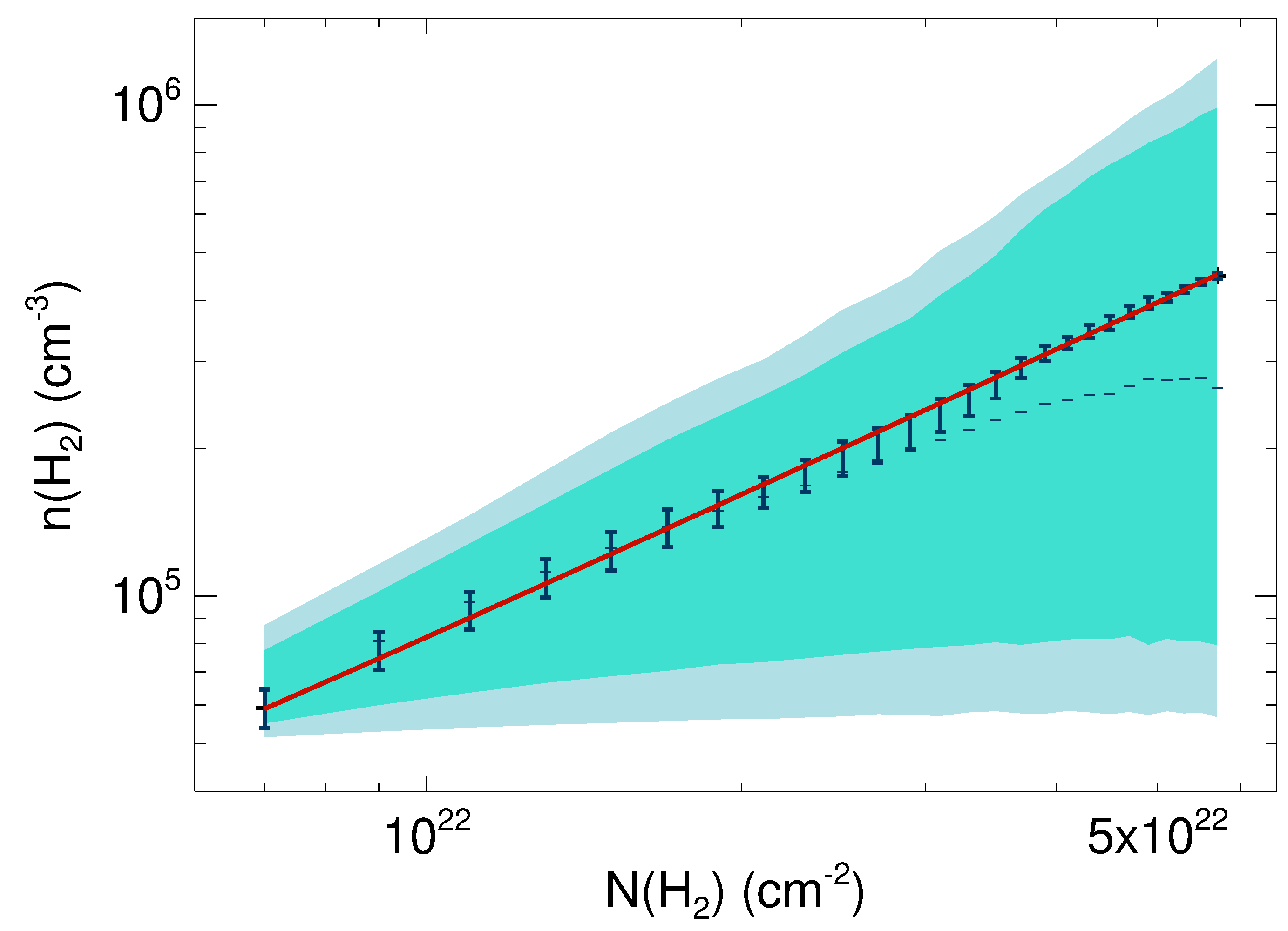}
    \includegraphics[height=6cm]{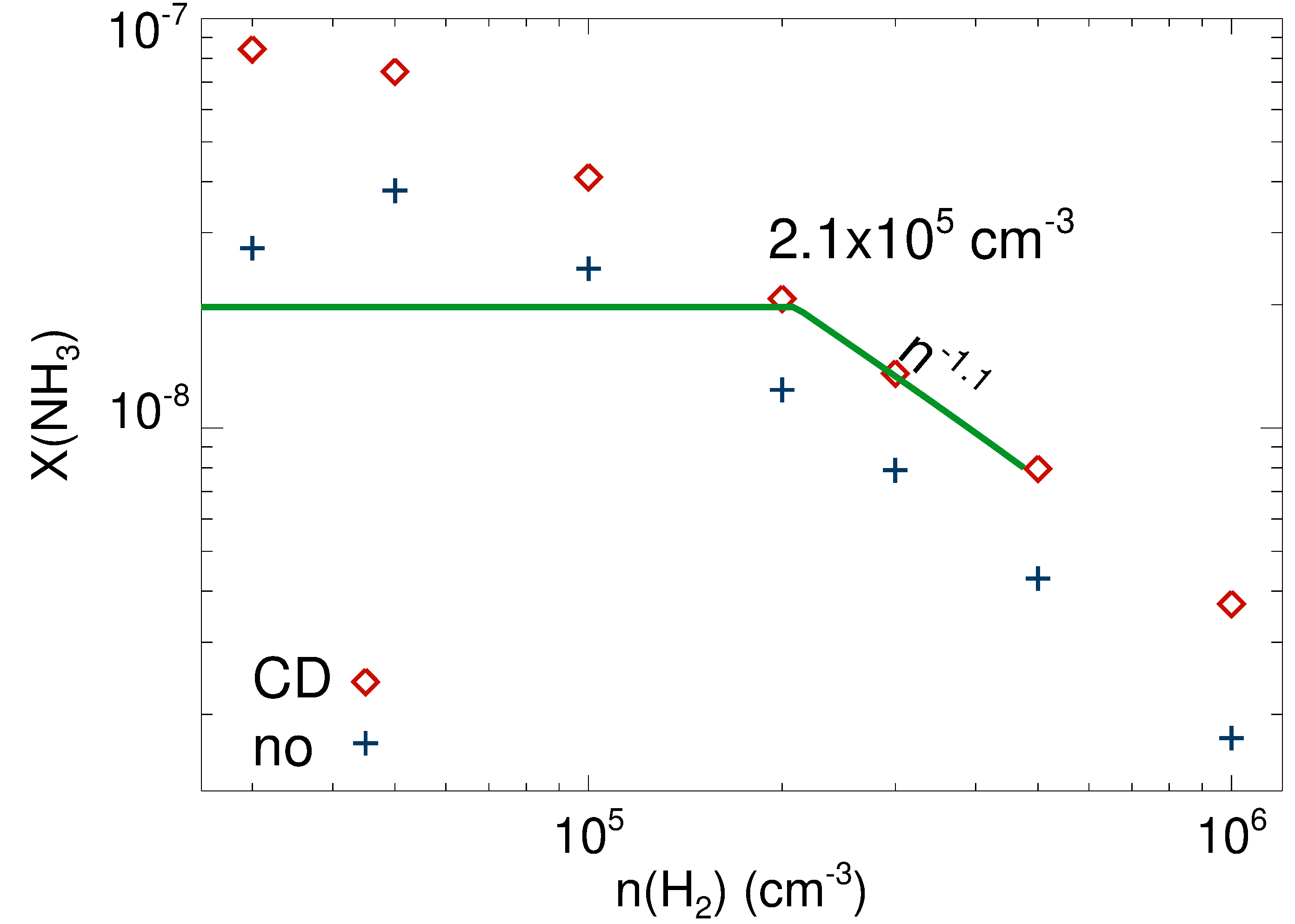}
    \caption{{\it Left:} Gas density, $n(\htwo)$, as a function of the
    $\htwo$ column density in a 3-D model of the H-MM1 core (see text).
    The shaded regions show the densities between the 5th and 95th percentile 
    (pale blue) and 16th and 84th percentile (turquoise) within a certain column
    density bin. The red line shows a linear fit to the weighted mean
    densities,  $\bar{n}(\htwo)$, where each line of sight (LOS) is weighted 
    according to standard deviation of the density ($1/\sigma_n^2$) along that
    LOS. The error bars correspond to the weighted sigmas of the sample densities. 
    The median density in each $N(\htwo)$ bin is {shown} with small 
    horizontal bars. 
    {\it Right:} Fractional $\ammo$ abundance as a function of gas
    density in H-MM1. The relationship is derived by fitting a broken
    line to mean $\ammo$ column densities below and above the turning
    point $N(\htwo)=2.6\times10^{22}\,\persqcm$ (see Sect.\,\ref{sec:X_calc}), and using a
    linear fit to $\bar{n}(\htwo)$ vs. $N(\htwo)$ on the left. Also shown
    are predictions from our chemistry model including chemical 
    desorption (CD, diamonds) and CD turned off 
    (plus signs; see text). 
    \label{fig:col2dens}}
\end{figure*}

The broken-line fit to the $N(\ammo)$ vs. $N(\htwo)$ 
{distribution} (Sect.~\ref{sec:X_calc})
and the inversion to mean densities described above, give the
relationship $X(\ammo) \propto n^{-1.1}$ at high densities. The
dependence of the fractional $\ammo$ abundance on the density is
shown in Fig.~\ref{fig:col2dens}-right. 
The density threshold adopted in the conversion from column densities to average volume densities affects the break-point density in this diagram.
{Applying no threshold density would shift the break point to $1.0\times10^5\,\percc$. 
However, the slope of the decreasing abundance
would remain the same. We note that the pivotal density as
well as the fractional $\ammo$ abundance at low densities are likely
to depend on the temperature, the external radiation field, and the
chemical age of the core, and they probably change from core to
core.} 

Also shown in {Fig.~\ref{fig:col2dens}-right} are data points
from a series of one-point chemistry models at different densities with the gas and dust temperatures fixed to 11\,K. The chemistry code is described in Sect.~\ref{ss:modeling}, and the simulations used to investigate the density dependence are described in Sect.~\ref{ss:density_dependence}. 

\begin{deluxetable}{cccc}
\tablecaption{Summary of Reported \ammo Abundances\label{tab:X_NH3}}
\tablewidth{0pt}
\tablehead{
\colhead{Source} & \colhead{Method} & \colhead{[\ammo/\htwo]} 
& \colhead{Ref.} 
}
\startdata
H-MM1 & Int. Col.Rat. & $2.7\times10^{-8}$ & This work\\
H-MM1 & Int. Linear Fit & $(1.975\pm0.005)\times10^{-8}$ & This work\\
\hline
H-MM1 & SD Col.Rat. & $(0.76\pm0.02)\times10^{-8}$ & 1\\
L1498 & SD Mod. & $3.4 \times10^{-8}$ & 2\\
L1517B& SD Mod. & $2.8 \times10^{-8}$ & 2\\
L1688 & SD Col.Rat. & (0.3--3)$\times10^{-8}$ & 3\\
L1544\tablenotemark{a} & Int. Mod. & (0.3--0.8)$ \times10^{-8}$ & 4\\
\enddata
\tablenotetext{a}{An abundance profile is reported.}
\tablerefs{
(1) \cite{Harju2017-HMM1_NH2D}, 
(2) \cite{Tafalla2004-Cores},
(3) \cite{Friesen2017}, 
(4) \cite{Crapsi2007-L1544}.
}
\tablecomments{
SD = Single Dish data used; 
Int. = Interferometric data used.;
Col.Rat. = Column Density Ratio used to determine fractional abundance;
Mod. = fractional abundance constrained via radiative transfer modelling.
}
\end{deluxetable}

\subsection{Chemistry of ammonia}
\label{ss:ammonia_chemistry}

The formation of ammonia in the interstellar gas is 
recapitulated in, for example, \cite{2015A&A...576A..99R} and
\cite{Sipila15b}. Once the \ce{NH+} ion is available, a sequence of
reactions with \htwo lead quickly to the ammonium ion, \ce{NH4+},
which is the principal precursor of \ammo. Electron
recombination of \ce{NH4+} and smaller cations give the
nitrogen hydrides \ce{NH}, \ce{NH2}  and \ammo.  Ammonia is
expected to be the most abundant of these in dense cores.
The pivotal ion, \ce{NH+}, is formed by the charge transfer reaction
\ce{NH + H+ -> NH+ + H}
or by the reaction 
\ce{N+ + ortho-H2 -> NH+ + H}.
In dark cloud 
conditions, the \ce{N+} ion forms by the dissociative charge 
transfer reaction 
\ce{N2 + He+ -> N+ + N + He}
\citep{2010A&A...513A..41H, 2020A&A...643A..76H}.

In the gas, nitrogen hydrides are constantly converted to
cations through charge or proton exchange reactions, and returned to
neutral species in recombination reactions with electrons.
The dominant ions reacting with ammonia are \ce{H+} and \ce{H3+}
with approximately equal destruction rates, and the ionic species
formed in these reactions, \ce{NH3+} and \ce{NH4+}, are quickly
returned to ammonia; \ce{NH3+} reacts with \htwo to form
\ce{NH4+}, which gives mainly \ammo in recombination with  electrons
(see Fig.~2 in \citealt{Sipila15b}).

Neutral nitrogen hydrides, as well as nitrogen atoms and 
molecular nitrogen, are also likely to be adsorbed onto dust
grains, and this should eventually diminish the abundances of
ammonia and related species, unless they are returned into the
gas to the same degree. The accreted \ce{N} atoms, and the \ce{NH} and 
\ce{NH2} radicals, are efficiently hydrogenated to \ammo,
and most nitrogen in interstellar ices is probably bound to
ammonia. As a consequence, ammonia is much more abundant in the
icy mantles of dust grains than in the gas phase, and the ice
mantles constitute the biggest ammonia reservoir in dense
molecular clouds. In quiescent, shielded regions, the most
important mechanisms releasing part of this ammonia probably 
are the cosmic-ray-induced desorption 
\citep[e.g.,][]{1985A&A...144..147L} and reactive desorption 
\citep[e.g.,][]{2006FaDi..133...51G}. At very high densities, where the accretion timescale is short, the gas-phase abundance of ammonia probably depends on the balance between accretion and desorption.

\subsection{Chemical Modeling}
\label{ss:modeling}

To aid the interpretation of the present observational results which
indicate strong depletion of ammonia, we simulated the chemistry of 
\ammo in H-MM1 using our rate-equation gas-grain astrochemical model
\citep[e.g.,][]{Sipila15b,Sipila19b}. In brief, the model solves a set of
rate equations connecting chemical reactions in the gas phase and on
grain surfaces. Several desorption mechanisms are considered: thermal
desorption, cosmic-ray induced desorption, photodesorption, and chemical
desorption. We employ a one-dimensional physical model that was
constructed by extracting density and dust temperature cuts through the
density peak of the three-dimensional model described in
Sect.\,\ref{ss:depletion} and making these cuts symmetric with respect to
the density peak by averaging. The dust temperature distribution of the
three-dimensional model was calculated by exposing the density model to
an external radiation field composed of a) diffuse isotropic
interstellar radiation, and b) radiation from B-type stars located 
$\sim$1\,pc west of the core. The intensities of these components were
adjusted until the 850\,$\mum$ surface brightness map from the model
agreed with the SCUBA-2 map and the average dust temperature map agreed
with the color temperature $T_{\rm C}$ map derived from {\it Herschel}. We 
discretized the physical model into a series of points, and ran the
chemical simulations separately in each point to obtain time- and 
radially-dependent \ammo abundance profiles. We employ here the same
initial chemical abundances and other chemical model parameters as in
\citet{Harju2017-HMM1_NH2D}. The physical model is shown in
Fig.\,\ref{fig:HMM1_physicalModel}, and the initial abundances 
are listed in Table~\ref{tab:initialabundances}.

\begin{table}[t]
	\centering
	\caption{Initial chemical abundances (with respect to $n_{\rm H}$).}
	\begin{tabular}{cc}
		\hline
		\hline
		Species & Abundance\\
		\hline
		$\rm H_2$ & $5.00\times10^{-1} \,^{(a)}$\\
		$\rm He$ & $9.00\times10^{-2}$\\
		$\rm C^+$ & $7.30\times10^{-5}$\\
		$\rm N$ & $5.30\times10^{-5}$\\
		$\rm O$ & $1.76\times10^{-4}$\\
		$\rm S^+$ & $8.00\times10^{-8}$\\
		$\rm Si^+$ & $8.00\times10^{-9}$\\
		$\rm Na^+$ & $2.00\times10^{-9}$\\
		$\rm Mg^+$ & $7.00\times10^{-9}$\\
		$\rm Fe^+$ & $3.00\times10^{-9}$\\
		$\rm P^+$ & $2.00\times10^{-10}$\\
		$\rm Cl^+$ & $1.00\times10^{-9}$\\
		\hline
	\end{tabular}
	\tablenotetext{a}{The initial $\rm H_2$ ortho/para ratio is $1 \times 10^{-3}$.}
	\label{tab:initialabundances}
\end{table}

\begin{figure}
\centering
        \includegraphics[width=1.0\columnwidth]{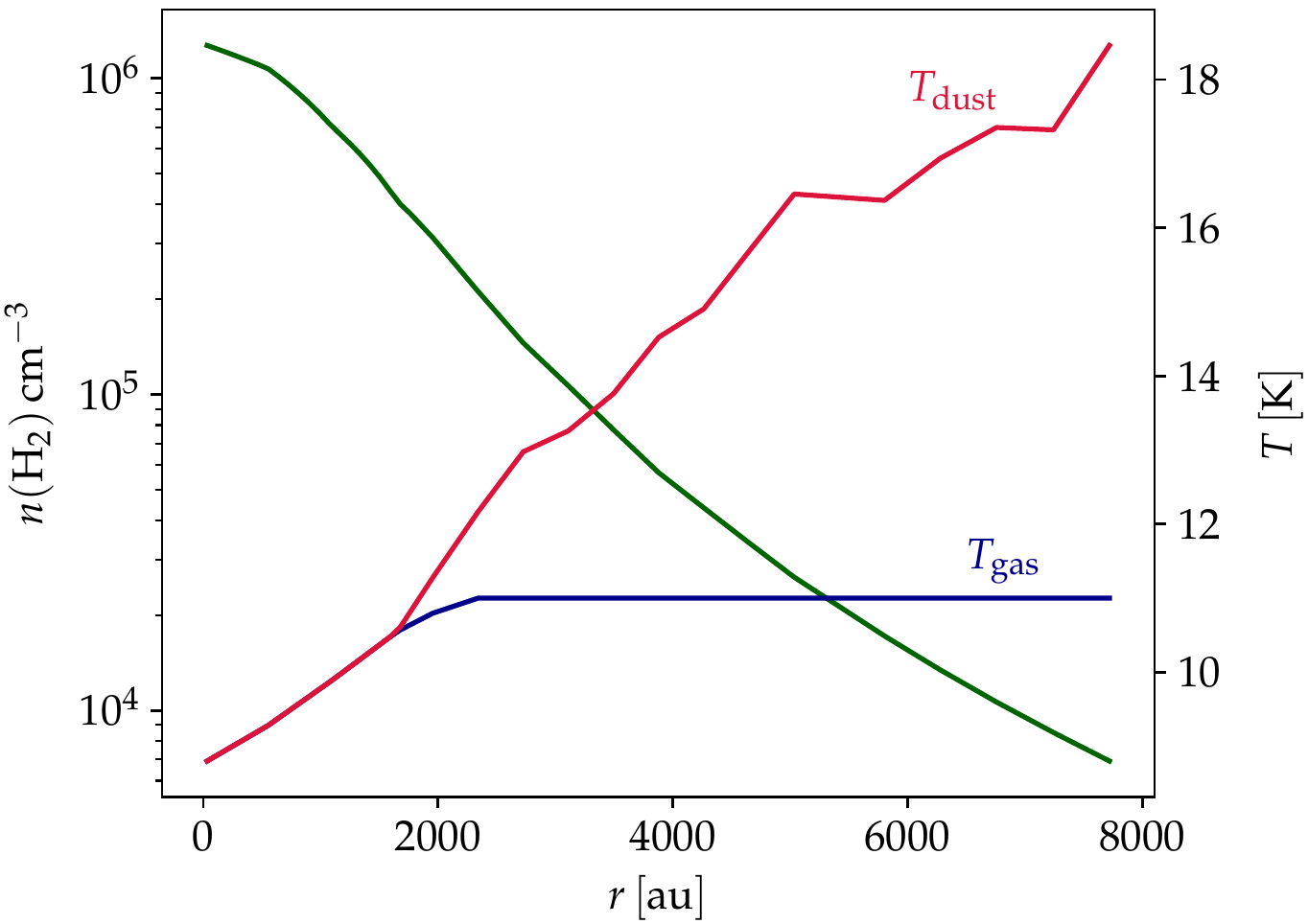}
    \caption{Physical model of H-MM1 adopted in chemical simulation: 
    \htwo number density (green) as well as the gas (blue) and dust (red) 
    temperatures as functions of the radial distance from the center.}
    \label{fig:HMM1_physicalModel}
\end{figure}

Based on the derived abundance profiles, we then simulated the \ammo 
(1,1) and (2,2) lines using the radiative transfer code LOC \citep{Juvela20}. 
The line profiles and the column density were convolved to
the angular resolution of the observations (6$\arcsec$).
In \citet{Harju2017-HMM1_NH2D}, we derived a best-fit simulation time of
$3 \times 10^5 \, \rm yr$ based on fitting multiple deuterated ammonia
lines. For consistency we adopt the same time here.

\begin{figure*}
\centering
        \includegraphics[width=2.0\columnwidth]{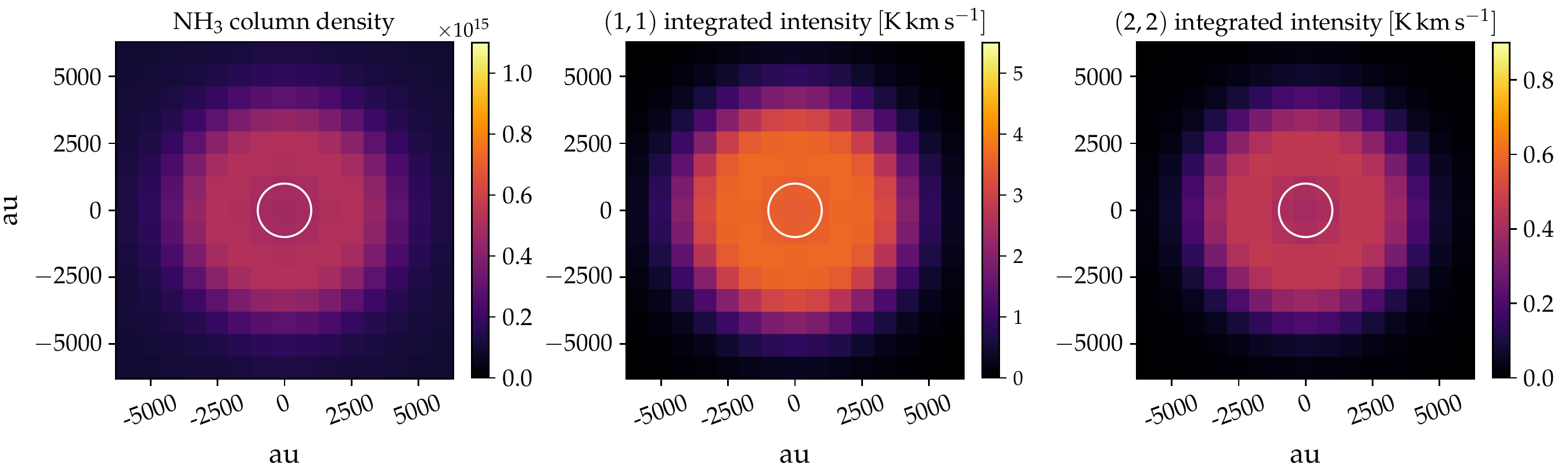}
    \caption{Simulated p\ammo column density (left), 
    and integrated intensity ((1,1) and (2,2); middle and right panels, respectively) maps. 
    The white circle in each panel indicates a region of 2000\,au diameter.}
    \label{fig:HMM1_model}
\end{figure*}

Figure~\ref{fig:HMM1_model} shows the total \ammo column density
(summed over ortho and para \ammo) and integrated intensity maps
obtained at this time step. The plots are displayed using the same
scaling for the color bar as in
Figs.\,\ref{fig:NH3_TdV} and \ref{fig:column_maps}, and we have added
$10^{14} \,\persqcm$ to the simulated column density to account for
the ammonia in the outer cloud that is not included in the core model
(see Sect.\,\ref{ss:depletion}). The match between the model
and the observations is quite good in the innermost region of the maps,
which represents the volume immediately surrounding the \htwo column
density peak in H-MM1, even if the model somewhat under-produces the
emission in both the (1,1) and the (2,2). Although not very clearly
visible in the plot, both the column density and the integrated intensity
maps present a dip toward the center of the core. Naturally, the model
does not recover the morphology of the observed column density or
integrated intensity maps outside the central region as we are using a
spherically symmetric 1D model. The strength of the line emission is
sensitive to uncertainties in the gas temperature in the outer core where
the lines are produced.

The fractional \ammo abundance is higher near the western edge of
the core than at the eastern edge (Fig.~\ref{fig:NH3_abundance}-right). 
This asymmetry can
probably be attributed to the fact that the temperature gradient is
steeper on the western side owing to intense radiation from that
direction. We examined this hypothesis by using two one-dimensional
models, which we call ``east'' and ``west'', with the density and
temperature profiles corresponding to the eastern and western sides of
the core (instead of a symmetrized profile used above). The steeper
rise of the temperature as a function of the distance from the core
center in the ``west'' model was also accompanied with a steeper
increase of the \ammo abundance as compared to the ``east'' model.

\subsection{Density dependence of the ammonia abundance}
\label{ss:density_dependence}

As discussed in Section~\ref{ss:ammonia_chemistry}, the dominant
gas-phase reactions involving ammonia produce closely related molecular
ions that are quickly returned back to \ammo. Therefore, the diminishing
ammonia abundance at high densities must ultimately be caused by
adsorption of \ammo and related species, mainly \ce{NH}, \ce{N}, and \ce{N2}, 
onto grains. The rate at which an atom {or} molecule accretes onto grains
is proportional to its average thermal speed and the total surface 
area of dust grains.  The latter can be expressed as the
product of the grain surface area per H atom, $\sigma_{\rm H}$ , and
the total hydrogen density, $n_{\rm H}$ ($\sim 2\times n(\htwo)$). 
{In other words, the accretion rate is proportional to the density.}

We examined the dependence of $X(\ammo)$ and related species on
$n(\htwo)$ by employing our full chemical network. This was done by
running a series of one-point models at different densities with the
gas and dust temperatures fixed to 11\,K. Ammonia abundances were
extracted at the simulation time $3\times10^5$\,yr. Two different
set-ups were used: one with chemical desorption (CD) and another with
CD turned off. The results are shown in Fig.~\ref{fig:model_x_vs_n}. 
One can see that at densities above $\sim 10^5\,\percc$, $X(\ammo)$
follows the decreasing abundances of \ce{N} and \ce{N2}. The slope of the
$\log X(\ammo)$ vs. $\log n(\htwo)$ curve is $-1.04\pm0.01$ in the model
with CD, and a little steeper, $\sim -1.15\pm0.04$, when CD is turned off. 
The dominance of accretion leads to a density
dependence close to $n^{-1}$ for neutral species, whereas 
electrons and light ions (\ce{H+} and \ce{H3+}) show a shallower
power law, $\sim n^{-0.44}$ in our models. The latter is determined by
the competition between the cosmic ray ionization and neutralization in 
recombination reactions.  

\begin{figure*}[htb]
    \centering
    \includegraphics[height=6cm]{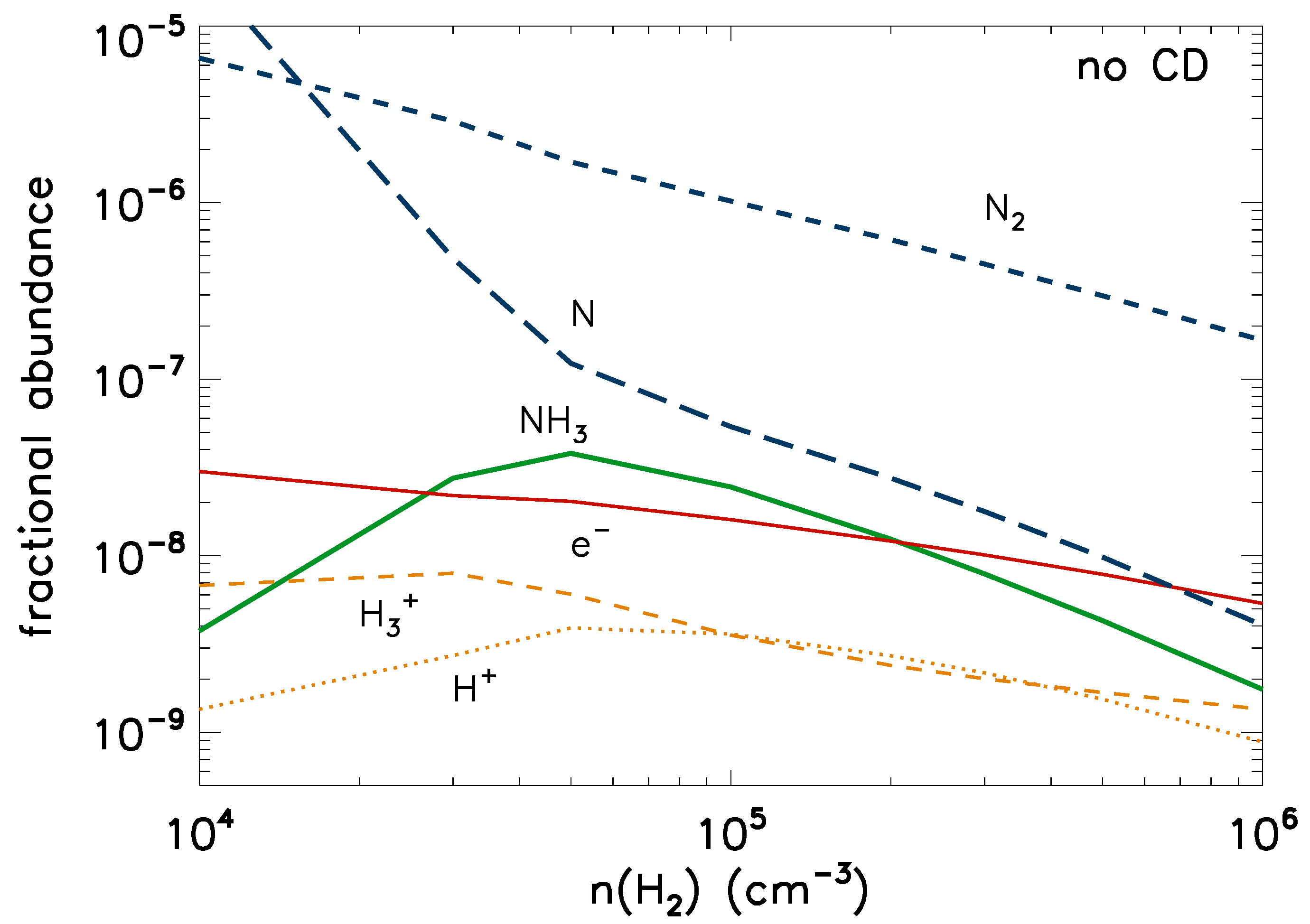}
    \includegraphics[height=6cm]{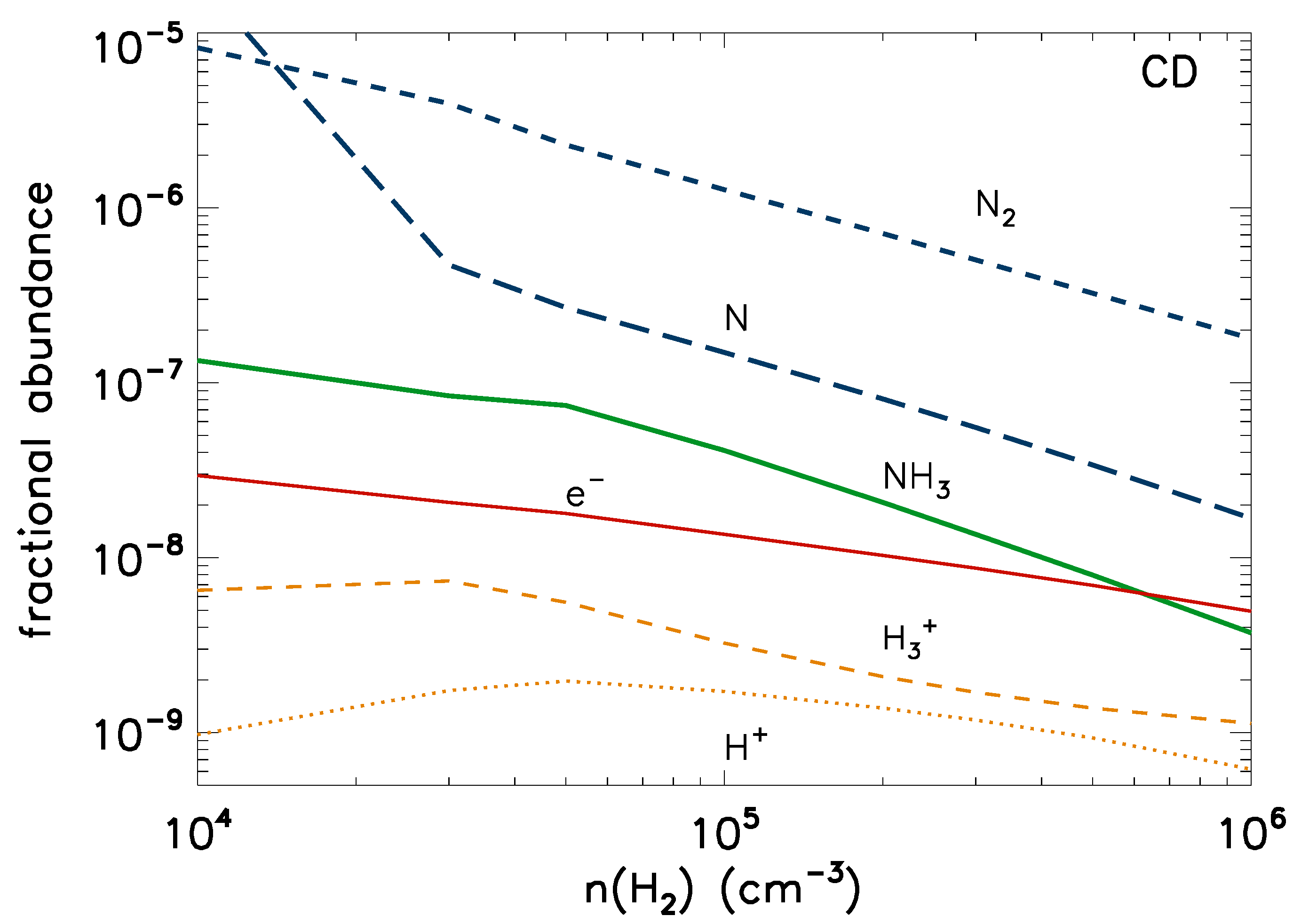}
    \caption{Modeled fractional abundances of selected species as functions of the volume density at the time $3\times10^5$\,yr after the beginning of the simulation. 
    {\it Left:} A model without chemical desorption (CD). 
    {\it Right:}  A model with CD turned on, assuming that 1\% of exothermic surface reactions will lead to the evaporation of the product. 
    \label{fig:model_x_vs_n}}
\end{figure*}

According to our model, most of ammonia is to be found on dust grains, where it reaches a
fractional abundance of $\sim10^{-4}$ relative to $\htwo$, and stays
almost constant to the end of the simulation.  
Replenishment of the
gas with ammonia from grains is significant when CD is
effective. Consequently, $X(\ammo)$ decreases more slowly and at higher
densities in the model with CD than in that without CD. 

Inspection of reaction rates during our chemical simulation shows that
the density $5\times10^{4}\,\percc$, where the \ammo abundance turns
down, is where the adsorption rate of ammonia exceeds the production
of \ce{NH+} ions in the gas. In the model with CD, desorption from
grains always gives more ammonia than ion-molecule reactions, but the
adsorption rate surpasses the desorption rate at a density of
$\sim 1\times10^{5}\,\percc$. At still higher densities, above
$n(\htwo)=2\times10^{5}\,\percc$, a balance between the primary
production of gaseous ammonia and its adsorption is established: 
the adsorption rate is equal to the combined rates of desorption and
\ce{NH+} production up to the highest density modeled, $10^6\,\percc$.
In our CD model, desorption accounts for $\sim 80\%$ of
ammonia production in this density range. 

The dominant surface reaction releasing ammonia into the gas in this
model is 
\ce{NH2^* + H^* -> NH3}.
Here we have indicated
species residing on the surface with asterisks. The mentioned reaction is the
end point of hydrogenation of N atoms trapped on grains. On the other
hand, a majority of \ce{NH2^*} molecules on the surface is formed by
dissociation of $\ammostar$, either  by cosmic rays or by secondary photons
induced by these.  We can therefore present a simplified picture, where 
desorption is proportional to the density of ammonia residing on grains
and a desorption coefficient. This picture is also valid for direct
ammonia desorption caused by external agents, such as grain heating
through cosmic-ray hits.

{The equilibrium condition between adsorption and desorption
can be written as $k_{\rm ads}\, n(\ammo) = k_{\rm des}\, n(\ammostar)$,
where $k_{\rm ads}= n_{\rm H} \sigma_{\rm H} {\bar v}_{\ammo}$ is the
adsorption rate per molecule ($\pers$), ${\bar v}_{\ammo}$ is the
mean thermal speed of ammonia molecules, and $k_{\rm des}$ is the
desorption rate per molecule ($\pers$). The ammonia densities in the
gas and on grains (per cubic centimeter) have been denoted by $n(\ammo)$ and
$n(\ammostar)$. Dividing the equilibrium equation by $n(\htwo)$ and
re-arranging the terms one gets for the fractional ammonia abundance
$X(\ammo) = k_{\textrm{des}} \, X(\ammostar)/(\sigma_{\rm H}\, {\bar v}_{\ammo}
\,n_{\rm H})$. According to our chemistry model, the ammonia abundance in ice 
exceeds the gas-phase abundance by several orders of magnitude in starless dense cores, 
and accretion of ammonia from the gas does not change $X(\ammostar)$ significantly 
at this stage. Assuming that also the desorption rate is approximately constant, 
balance between adsorption and desorption leads to an $n^{-1}$ dependence of
$X(\ammo)$ in the gas. }

{As can be seen from Fig.~\ref{fig:col2dens}-right, both} 
model set-ups, with or without CD, fail to reproduce
the observed abundances at low densities. This probably depends on
the fact that the simulation starts from atomic gas composition in
physical conditions appropriate for a dense core. In reality, the
dense core material has already been processed in ambient molecular
cloud conditions with $n(\htwo)\sim 10^3-10^4\,\percc$, where ammonia
production on grains and in the gas has already started, and a 
substantial fraction of nitrogen is already incorporated in ammonia ice. 

\section{Conclusions}

We present high spatial resolution ammonia observations with the Very Large Array and 
Green Bank Telescope 
toward the starless core H-MM1 in Ophiuchus.
They indicate that ammonia has
a constant fractional abundance, $X(\ammo)={(}1.975{\pm0.005)}\times 10^{-8}$ 
{($\pm$10\% systematic)}, 
up to a column density of $N(\htwo) = 2.575\times 10^{22}\,\persqcm$,
beyond which the abundance starts to decrease.
This is direct evidence of the \ammo depletion from the gas phase, 
in a region of size comparable to the single dish beam size, which 
explains why the depletion region is only detected 
in deep high-angular resolution observations.
By
deriving a volume density model of the core we established that the
critical density where \ammo starts to disappear from the gas is
approximately $2\times10^5\,\percc$ in this core, and at higher
densities $X(\ammo)$ follows the power-law $X(\ammo)\propto n^{-1.1}$. This
tendency at high densities is reproduced by chemical simulations, and
is consistent with the idea that adsorption onto grains is the main
mechanism {reducing the gas-phase abundances} of neutral molecules 
in the core center.  Our
chemistry simulations suggest that the break-point in $X(\ammo)$
versus $n(\htwo)$ relationship marks the density where the adsorption
rate surpasses the production rate of gaseous ammonia through
ion-molecule reactions and desorption.  The density threshold for
depletion is likely to vary from core to core because it depends on
the temperature and the total grain surface area per hydrogen atom.

\acknowledgments
We thank the referee for their thoughtful comments which 
improved the paper.
JEP, JH, PC, OS, MJM, SP, and DMS acknowledge the support by the 
Max Planck Society. 
DMS is supported by an NSF Astronomy and Astrophysics Postdoctoral Fellowship under award AST-2102405.
%

\facilities{VLA, GBT} 

\software{Aplpy \citep{aplpy,aplpy2019}, 
Astropy \citep{Robitaille_2013,Astropy2018}, 
Matplotlib \citep{Hunter_2007},
SciPy \citep{2020SciPy-NMeth},
pyspeckit \citep{pyspeckit},
spectral-cube \citep{spectral-cube},
EMCEE \citep{EMCEE},
corner  \citep{corner}}

\bibliographystyle{aasjournal}
\bibliography{bibliography}

\appendix

\section{Possible effects of scattering and emission at $8\,\mum$ on the $N(\htwo)$ map \label{sec:scatter}}

\citet{Lefevre2016} discussed the possible effect of dust scattering
for the 8\,$\mu$m absorption. With models of the cloud LDN~183, they
concluded that scattering could correspond to several times
$0.1\,\MJysr$, helping to conciliate column density estimates
from mid-infrared (MIR) absorption and molecular line
emission. Scattering at this level required aggregate grains with
sizes much larger than in the general interstellar medium.

We examined H-MM1 similarly {with} 3D radiative transfer
models. The models were optimised to reproduce the {\it Herschel}
250--500\,$\mu$m observations under the assumption that the
line-of-sight density profile is similar to that derived from the
surface brightness, along a constant-declination cross section through
the core centre. The external radiation field was set according to the
\citet{Mathis1983} model, adopting the angular distribution on the sky
from COBE/DIRBE observations \citep{Hauser1998}. We further added an
energetically equal component for radiation from the equatorial west,
to qualitatively account for the effect of the nearby B-type star.
The optimisation of the model resulted in a scaling of the radiation
field and the column-density for each map pixel. The models reproduce
the {\it Herschel} surface brightness observations within 1\,\% at
350\,$\mu$m and within $\sim 10$\,\% at 250\,$\mu$m and
500\,$\mu$m. The model optimisation was repeated for several dust
models, including the {\cite{Compiegne2011}} dust model, the DustEM
core-mantle-mantle (CMM) and ice coated aggregates (AMMI) grains
{\citep{Ysard2016}}, and the \citet{WeingartnerDraine2001} $R_{\rm
V}=5.5$ ``B'' model. The estimated radiation fields were consistently
slightly more than ten times the \cite{Mathis1983} values but,
because of different far-infrared emissivities, the column densities
varied by a factor of $\sim$3 between the dust models. 
In each case,
we also computed 8\,$\mu$m maps for the scattered light and for the
emission due to stochastically heated grains. All calculations were
performed with the radiative transfer program SOC \citep{Juvela2019}.

For all of the tested dust models, the 8\,$\mu$m scattered light
remained below $0.1\,\MJysr$. The largest value was obtained for
the \citet{WeingartnerDraine2001} grains, some $0.055\,\MJysr$,
the AMMI grains produced less than half of this value. The CMM and
AMMI models do not contain small grains and therefore result in no
significant MIR emission. The level of scattered light depends not
only on the scattering properties of the grains but equally on their
far-infrared emissivity that, via the fits to {\it Herschel} data, determines
the model column densities. The results suggest that (in the absence
of very large aggregates) the scattering is not a significant
factor. The 8\,$\mu$m extended sky brightness in H-MM1 is of the
order of $10\,\MJysr$, much higher than for LDN~183. Therefore,
more than to the scattered light, the analysis of MIR extinction is
sensitive to the assumed fraction of the extended sky brightness that
originates behind the core.

For the \cite{Compiegne2011} and \cite{WeingartnerDraine2001} dust
models, 
the emission from stochastically heated grains at 8\,$\mu$m
reaches a level of a few $\MJysr$. However, the
surface-brightness difference between the inner core and its
surroundings was observed to be either positive or negative, depending
on how the line-of-sight extent of the cloud was assumed to be
correlated with the column density. If the abundance of small grains
decreases towards the core centre, their emission is more likely to
bias the MIR extinction estimates upwards than downwards.

\section{Comparison of \ammo excitation and kinetic temperatures}
\label{sec:Tex_Tk_H2}
{
In the case of LTE, the excitation temperature and gas 
temperature are the same. Our determination of both these 
quantities allow us to directly assess how close to LTE 
the \ammo emission is.
The comparison of the excitation to kinetic temperature ratio 
as a function of \htwo column density is shown in
Fig.~\ref{fig:Tex_Tk_H2}. It shows that below 
$2\times10^{22}$ \persqcm the ratio increases with column density,
and at higher column densities
the excitation temperature is $\approx$90\% of the 
kinetic temperature value and close to LTE.
\begin{figure}[h]
    \centering
    \includegraphics[width=0.5\textwidth]{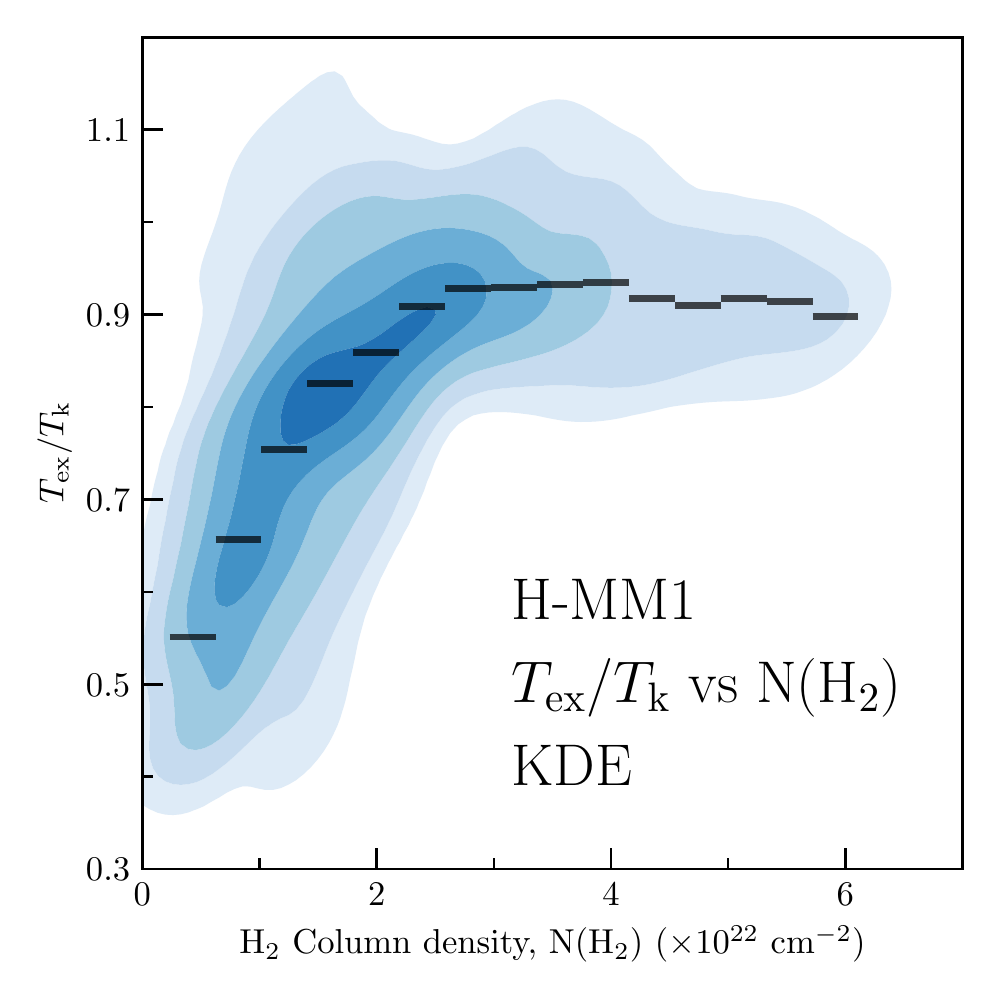}
    \caption{\ammo emission in LTE in the center of H-MM1.
    KDE comparison between the ratio of the excitation and 
    kinetic temperature derived from \ammo 
    and \htwo column densities. 
    In the case of perfect LTE the temperature ratio is 1. 
    The data, 
    whereas for higher $N(\htwo)$ values the temperature 
    ratio is nearly constant {$\approx$0.9, displays a behaviour very close to LTE}. 
    The black horizontal segments show the mean values within the 
    $N(\htwo)$ bins.}
    \label{fig:Tex_Tk_H2}
\end{figure}}

\section{Bayesian line fit \label{sec:X_fit}}
We  use \verb+emcee+ \citep{EMCEE} to fit a broken straight line to the 
\ammo and \htwo column densities. 
We use the estimated errors of $N(\ammo)$ derived from the \verb+pyspeckit+ fit,
an uninformative (uniform) prior to the variables to fit, and we run 50 chains for 15\,000 steps each.
The autocorrelation time is between 44 and 46 steps for the different parameters, and we
use the posterior distribution after using a burn-in of 150 steps 
and thinning the chains by a factor of 25.
The corner plot of the marginalized posterior distributions of the parameters is shown in Figure~\ref{fig:corner_plot}.

\begin{figure}
    \centering
    \includegraphics[width=0.9\textwidth]{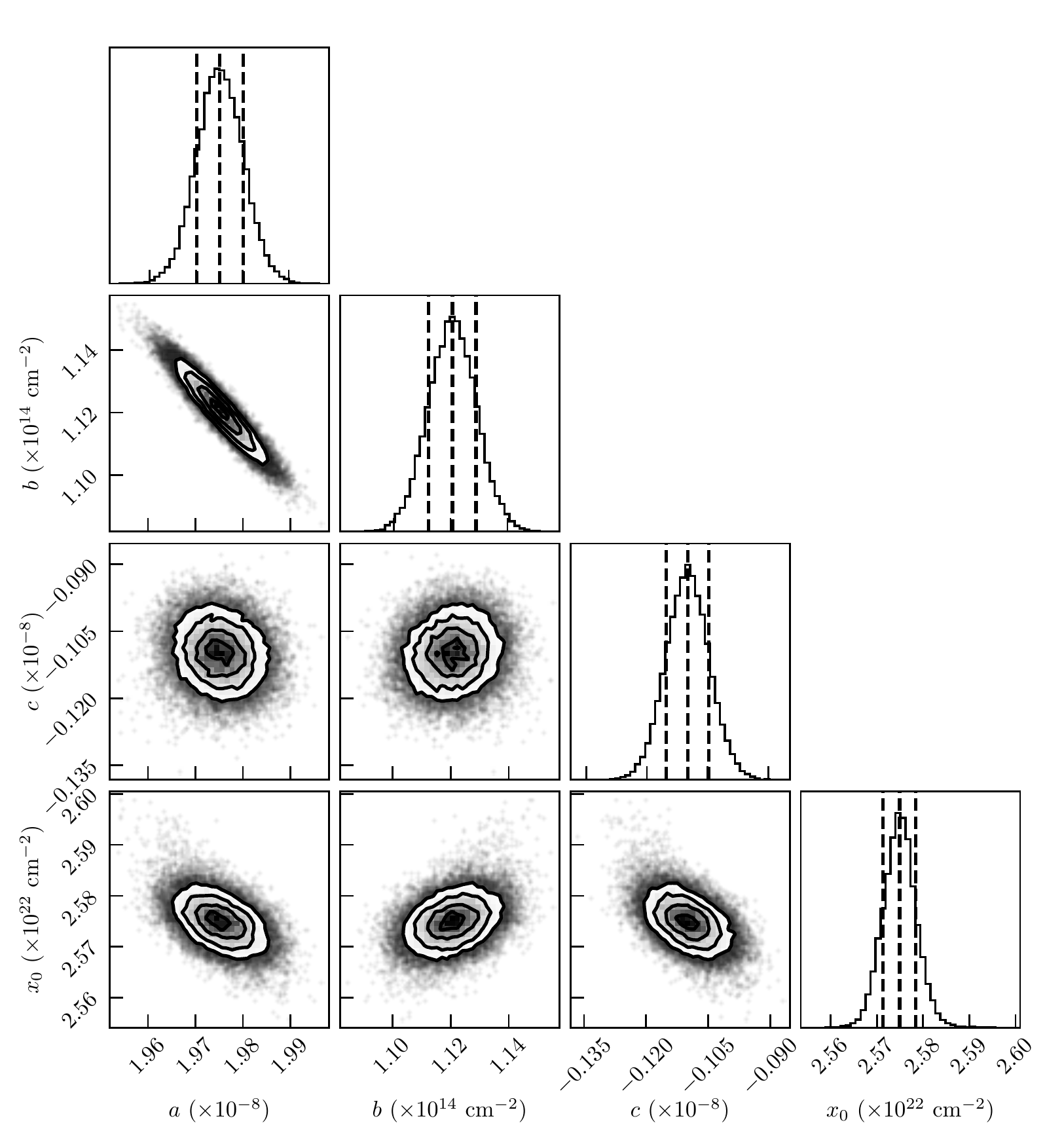}
    \caption{Corner plot of the marginalized posterior distributions for each parameter.
    The 16, 50, and 84 percentile of each marginalized posterior
    distributions are shown by the vertical dashed lines.
    The results are summarized in Table~\ref{tab:X_NH3_fit},
    where $a$ and $b$ are the $[\ammo]/[\htwo]$ abundance 
    (in units of $10^{-8}$) and 
    the \htwo column density offset (in units of $10^{14}\, \sqcm$)
    of the linear relation 
    in the lower column density region, respectively; 
    $x_0$ is the \htwo column density (in units of $10^{22}\, \sqcm$) 
    beyond which the depletion region is identified, 
    and $c$ is the $[\ammo]/[\htwo]$ abundance (in units of $10^{-8}$) 
    in the depletion region.}
    \label{fig:corner_plot}
\end{figure}

\section{Density model of H-MM1 \label{sec:density_model}}

{We estimate the density structure of the core by fitting a
Plummer-type function to the cross-sectional \htwo column density
profiles in different positions along the north-south oriented ridge
of the core.  The method is adopted from \cite{2011A&A...529L...6A}.
We assume that the density distribution has circular symmetry in the
plane perpendicular to the spine of the core. The latter is defined
by a spline fit through local $N(\rm H_2)$ peaks. The spline fit is
shown in Figure~{\ref{fig:density_fit}}a. A horizontal
(constant-declination) cut {through} the $N(\rm H_2)$ maximum of the core is
shown in Figure~{\ref{fig:density_fit}}b together with the fitted
function. The function is of the form 
$f(\Delta y)=\frac{a}{\left[1 + (\Delta y/b)^2\right]^c}$, 
where $\Delta y$ is the projected distance from the peak.  
The radial density profile,
$n(r)=\frac{n_0}{\left[1+(r/r_0)^2\right]^\eta}$,
is obtained from
this by the substitutions $\eta=c+0.5$, $r_0=b$, 
$n_0=a/(A_p r_0)$, where $n_0$ is the
central density and $A_p= \int_{-\infty}^{+\infty}
{(1+u^2)^{-\eta}} {du}$. For the cut shown in the figure, the
solution is $\eta=1.6\pm0.3$, $r_0=(1600\pm300)$\,au,
$n_0=(1.30\pm0.08)\times10^6\,{\rm cm}^{-3}$. The three-dimensional
density model is illustrated in Figure~{\ref{fig:density_fit}}c. Here,
the line of sight coincides with the negative {$x$-}axis, and the {$y$-}axis
points to the equatorial west. This density model is used for the mean-density vs. column density correlation shown in Fig.~\ref{fig:col2dens}-left.}

\begin{figure}[htb]
\unitlength=1mm
\begin{picture}(160,50)(0,0)
\put(-5,0){
\begin{picture}(0,0) 
\includegraphics[width=5.5cm,angle=0]{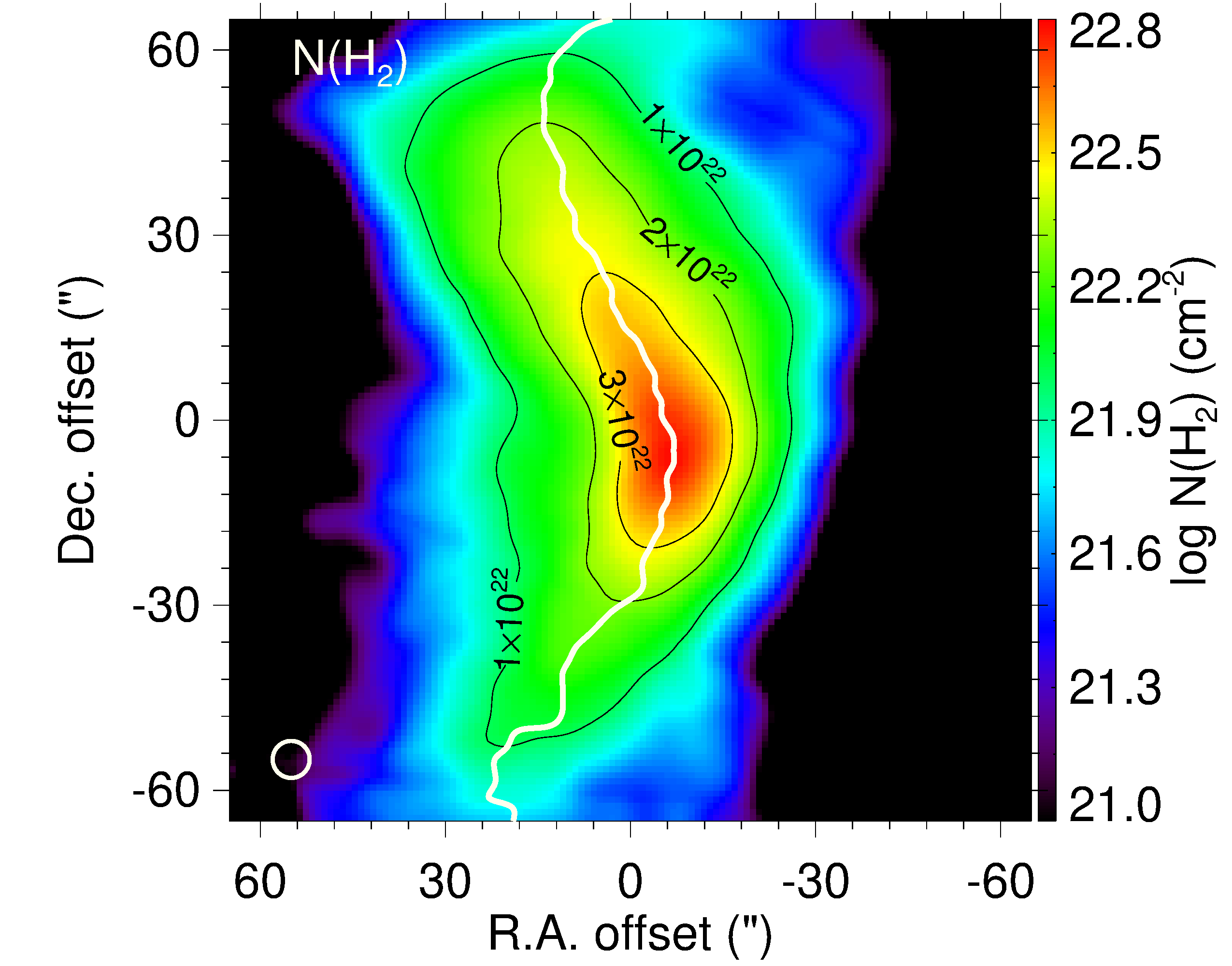}
\end{picture}}
\put(55,0){
\begin{picture}(0,0) 
\includegraphics[width=5cm,height=4.2cm,angle=0]{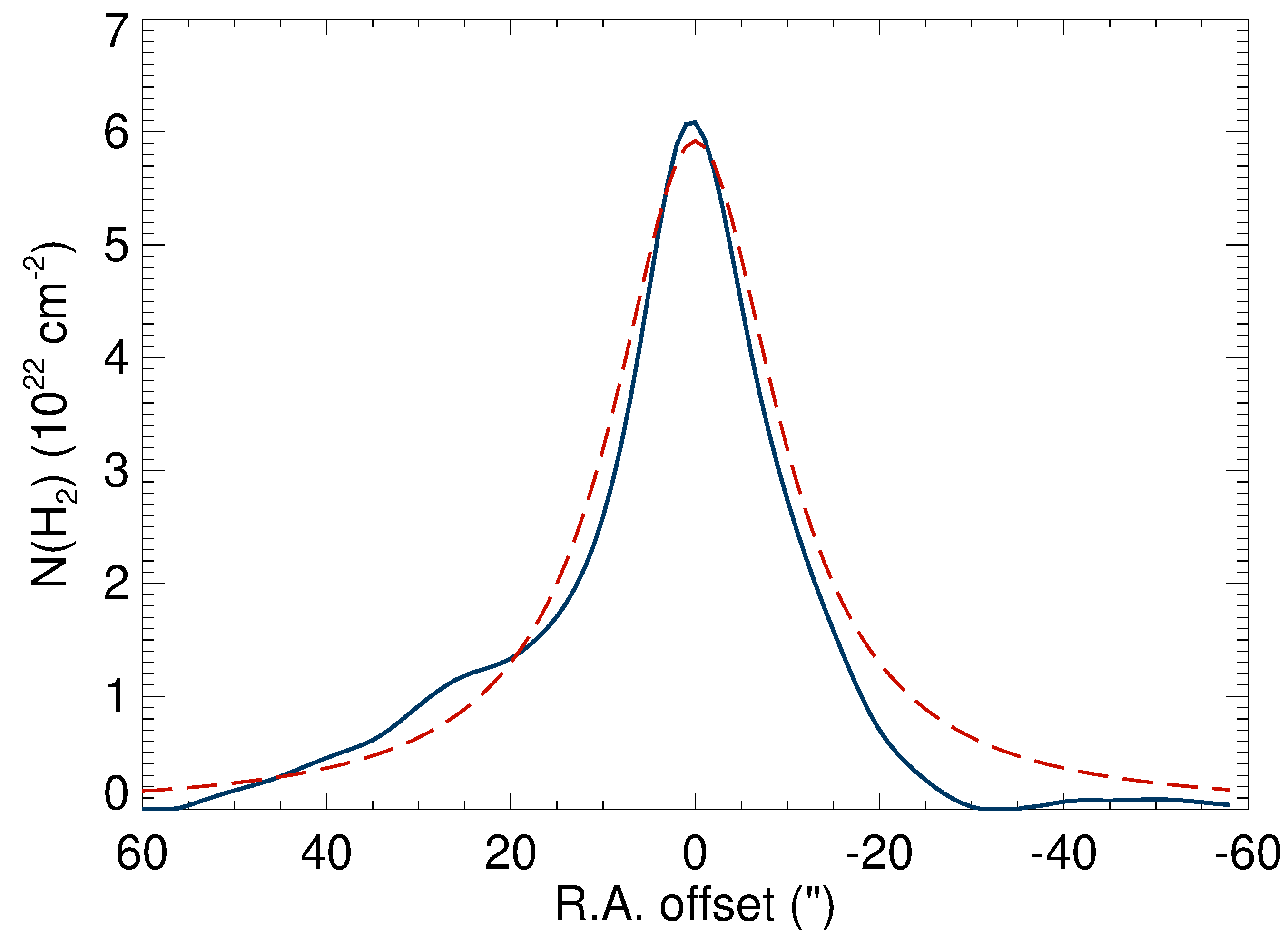}
\end{picture}}
\put(110,0){
\begin{picture}(0,0) 
\includegraphics[width=5cm,angle=0]{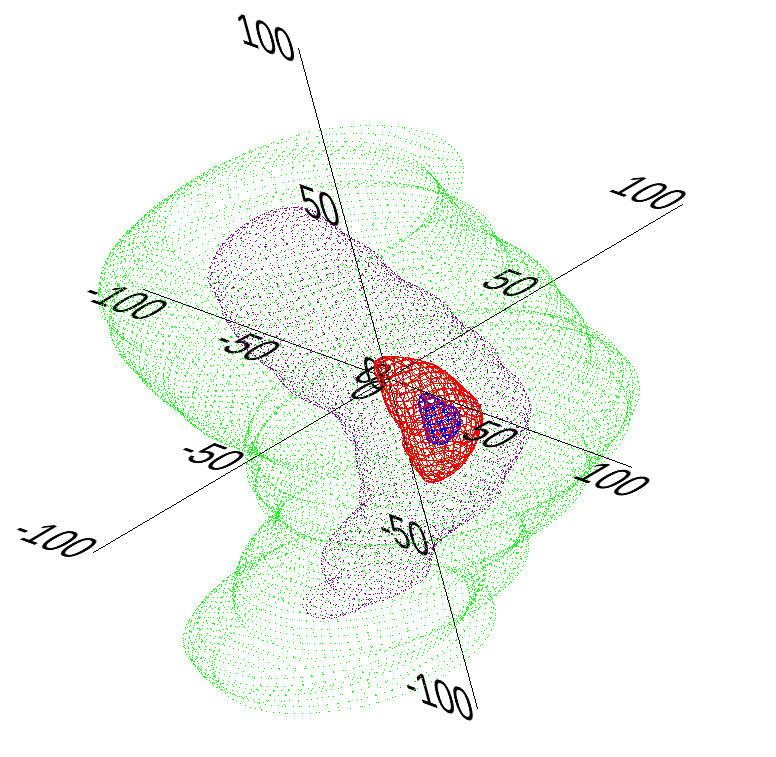}
\end{picture}}
\put(154,14){\bf x (toward us)}
\put(158,35){\bf y (west)}
\put(133,47){\bf z (north)}
\put(5,45){\bf a)}
\put(60,45){\bf b)}
\put(115,45){\bf c)}
\end{picture}
\caption{
{\bf a)}~ Spline fit through local $N(\htwo)$ peaks along the
ridge of H-MM1. {\bf b)} Horizontal cut through the $N(\htwo)$
maximum of the core together with a Plummer-type fit to the profile.
{\bf c)}~ Density model of H-MM1 constructed from Plummer fits.
The iso-density surfaces shown are: 
$n({\htwo})=10^4\,\cc$ (green), 
$10^5\,\cc$ (purple), 
$5\times 10^5\,\cc$ (red), and 
$10^6\,\cc$ (blue). The numbers indicate angular
separations in arc seconds. The distance to the object is {138.4}\,pc, so
a separation of $100\arcsec$ corresponds to {13,840} au.
\label{fig:density_fit}}
\end{figure}

\end{document}